\documentclass[12pt]{iopart}
\usepackage{indentfirst}
\usepackage[utf8]{inputenc}
\expandafter\let\csname equation*\endcsname\relax
\expandafter\let\csname endequation*\endcsname\relax
\usepackage{amsmath,amssymb,amsthm}
\usepackage{mathabx,mathrsfs}
\usepackage{esint}
\usepackage[colorlinks,breaklinks,linkcolor=blue,anchorcolor=blue,citecolor=blue,urlcolor=blue]{hyperref}
\usepackage{booktabs}
\usepackage{graphicx}
\usepackage{dcolumn}
\usepackage{bm}
\usepackage{bbm}
\usepackage{cite}
\usepackage[sort&compress]{natbib}
\setcitestyle{numbers,square,comma}
\usepackage{bm}
\usepackage{subfigure}
\usepackage{epstopdf}
\usepackage{lipsum}

\begin{document}

\title[Topological extension including quantum jump]{Topological extension including quantum jump}

\author{Xiangyu Niu}

\address{Center for Quantum Sciences, School of Physics, Northeast Normal University, Changchun 130024, China.}
\ead{niuxy795@nenu.edu.cn}

\author{Junjie Wang}

\address{Center for Quantum Sciences, School of Physics, Northeast Normal University, Changchun 130024, China.}

\begin{abstract}
Non-Hermitian systems and the Lindblad form master equation have always been regarded as reliable tools in dissipative modeling. Intriguingly, existing literature 
often obtains an equivalent non-Hermitian Hamiltonian by neglecting the quantum jumping terms in the master equation. However, there lacks investigation into the effects of discarded terms as well as the unified connection between these two approaches.
In this study, we study the Su-Schrieffer-Heeger model with collective loss and gain from a topological perspective. When the system undergoes no quantum jump events, the corresponding shape matrix exhibits the same topological properties in contrast to the traditional non-Hermitian theory. Conversely, the occurrence of quantum jumps can result in a shift in the positions of the phase transition. 
Our study provides a qualitative analysis of the impact of quantum jumping terms and reveals their unique role in quantum systems.
\end{abstract}

%
%
%
%
%

\section{\label{sec:intro}Introduction}
Over the years, research on topological band theory for closed quantum systems has witnessed considerable success, profoundly influencing various domains of quantum physics and material science. However, since quantum systems will inevitably interact with the external environment, it is essential to extend the corresponding study into open quantum systems, which may also bring about novel features to the topology. Crucially, previous studies mainly consider a non-Hermitian (NH) Hamiltonian as a simplification or substitution in studying the topological properties \cite{yao2018edge,xue2022nonhermitiana,xu2020topological,yoshida2019nonhermitian,lee2016anomalous,li2022nonhermitian,bergholtz2021exceptionala,bacsi2021dynamics,chang2020entanglement,kunst2018biorthogonal,legal2023volumetoarea,lieu2018topological}. Therefore, unraveling the extensions of topological phases in open quantum systems will become an urgent task. Despite being in its early stage, this research area has garnered increasing interest and ignited in-depth discussions \cite{nie2021dissipative,gneiting2022unraveling,bardyn2013topology,dangel2018topological}. 

Open quantum systems provide an influential framework for describing the exchange of energy, coherence, and information between a system and its surrounding environment \cite{breuerTheoryOpenQuantum2002a}. Under very general assumptions, the dynamics is governed by the Lindblad master equation, which forms a completely positive and trace-preserving Markovian map \cite{gardiner2004quantum}. However, the calculation cost of tracking the evolution of the density operator or solving the Liouvillian operator \cite{minganti2019quantum} can be extremely complicated, as it scales with the square of its Hilbert space dimension. Not to mention, in most cases, only numerical results are feasible, especially for intricate many-body systems.

By contrast, NH systems offer a simpler description of dissipation and can, to some extent, serve as a substitute for the master equation. It is because, from a structural perspective, the NH dynamics aligning with the continuous non-unitary evolution components of the master equation. However, another essential component known as the ``quantum jumping terms'', which refers to the stochastic jumps that occur during the evolution of the wave function, is often set to zero \cite{yamamoto2019theory,xu2020topological,yoshida2019nonhermitian}. Nevertheless, these terms may be non-negligible in many cases, since they could induce interesting phenomena or differences \cite{niu2023effect,minganti2019quantum}. For instance, they may commit abrupt perturbance towards the wave function \cite{daley2014quantum}, or affect the spectra of Liouvillian operators \cite{minganti2019quantum,torres2014closedform,niu2023effect}. In addition, NH systems still have unclear issues such as probability conservation, calculation of physical quantity averages, steady states, and spectral instability \cite{ashida2020nonhermitiana}. Thus their use and verification require careful consideration.
Consequently, exploring similarities and differences between these descriptions is of broad interest \cite{song2019nonhermitiana,longhi2020unraveling,yoshida2020fate}.  Naturally, one may wonder how the quantum jumping terms will affect the topological properties. Here, we will address this issue and show how the quantum jumping terms generate effects via a specific model.

In this paper, we focus on a Su-Schrieffer-Heeger (SSH) lattice with collective loss and gain, such a dynamic can be presented by a Lindblad master equation. Notably, when all the jumping terms are neglected, the system is dynamically equivalent to a NH SSH model \cite{yao2018edge}, with a pair of asymmetric hopping terms in the unit cell \cite{gong2018topologicala}.
By analyzing its shape matrix, which has emerged as a powerful tool for studying topological extensions \cite{dangel2018topological,kawasaki2022topological}, we consider two scenarios: the jump-present case and jump-absent case. Intriguingly, the topology appears to be entirely encoded within the shape matrix. For the former case, despite utilizing distinct frameworks, it remains capable of faithfully predicting the topological phase, while simultaneously ensuring the validity of the bulk-edge correspondence.
The introduction of quantum jumping terms may break the system's intrinsic symmetry, and lead to a shift in the phase transition points. These insights reveal the unique roles of quantum jumps in topological systems, and indicate that they shall be carefully
examined. Our work paves the way for defining the topological phase in open systems, which is a challenge of current research, and closely integrates the perspectives of NH systems and open systems together.

This paper is organized as follows. In Sec. \ref{sec:model}, we present the example model of the entire paper and introduce the basic theory, third quantization, to derive the shape matrix. In Sec. \ref{sec:symmetry}, we classify the symmetries of the model. In Sec. \ref{sec:GBZ}, we computed the generalized Brillouin zone of the system from the perspectives of two dissipation modeling frameworks. And in Sec. \ref{sec:winding number}, we defined their corresponding topological invariants. We also conducted an analysis of the phase transition points and discussed the preservation of the bulk-boundary correspondence. Finally, in Sec. \ref{sec:conclusion}, we conclude our results.

\section{\label{sec:model}Model}

\begin{figure}[htbp]
\centering
\subfigure{
\label{Fig:model}
\includegraphics[width=0.8\textwidth]{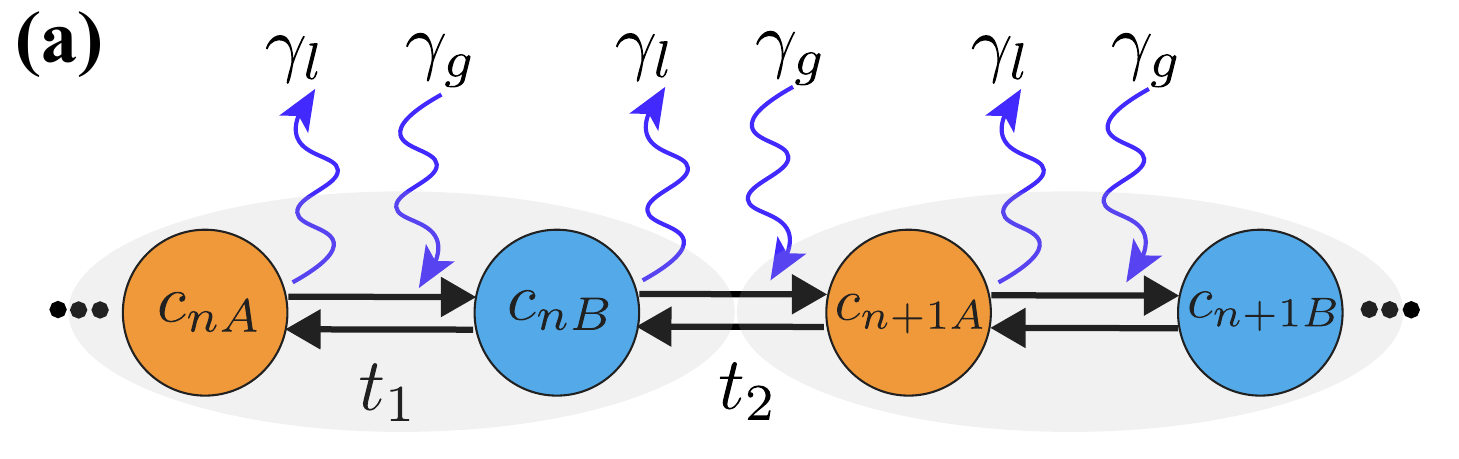}}
\subfigure{
\label{Fig:effmodel}
\includegraphics[width=0.8\textwidth]{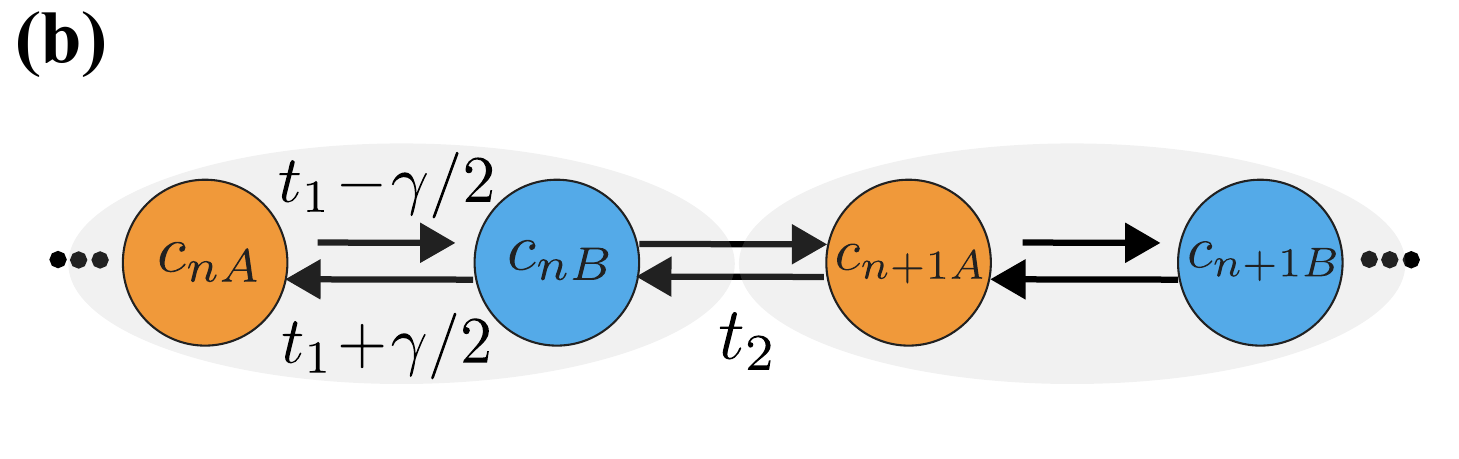}}
    \caption{(a) The $N$ site SSH model with inherent staggered hopping as well as loss and gain. The intracell hopping strength is $t_1$, and the intercell hopping is $t_2$. The strength of collective loss and gain is $\gamma_l$ and $\gamma_g$, respectively. When the quantum jump is absent, the dynamic is equivalent to (b), an effective non-reciprocal SSH model with hopping $t_1\pm\gamma/2$, when jumping in (a) is neglected, $\gamma=\gamma_l+\gamma_g$. }
\end{figure}

In this section, we start from a $N$ site one-dimension SSH lattice with dissipation, as shown in Fig. \ref{Fig:model}. The internal hopping strength between sites $A$ and sites $B$ is $t_1$, and the strength between the neighbor cells is $t_2$. The free Hamiltonian $H$ can be written as $H=\sum_{ij}h_{ij}c^\dagger_ic_j=\sum^{N/2}_n(t_1c^\dagger_{nA}c_{nB}+t_2c^\dagger_{n+1A}c_{nB}+\rm{H.c.})$, where $c$ is the annihilation operator of fermion and the subscript denotes the site $A,B$ in the $n^{\rm{th}}$ unit cell.

Considering the system is subject to loss and gain, it is convenient to be presented by the Lindblad master equation \cite{gardiner2004quantum}
\begin{equation}
    \dot{\rho}=\mathcal{L}\rho=-i\left[H,\rho\right]+\sum_{\mu}(2\kappa L_{\mu}\rho L^\dagger_{\mu}-\rho L^\dagger_{\mu} L_{\mu}-L^\dagger_{\mu} L_{\mu}\rho),\label{eq:LME}
\end{equation}
where $\mathcal{L}$ is the Liouvillian superoperator.

In the present model, the Lindblad operators $L_{\mu}$ include 
\begin{align}
        L_{l,n}=\sum_iD^l_{ni}c_i=\sqrt{\gamma_l/2}(c_{nA}+ic_{nB}),\label{eq:loss}\\
        L_{g,n}=\sum_iD^g_{ni}c^\dagger_i=\sqrt{\gamma_g/2}(c^\dagger_{nA}+ic^\dagger_{nB}),\label{eq:gain}
\end{align}
where $L_{l,n}$ stands for the collective loss and $L_{g,n}$ describes the gain, the $\gamma_l,\gamma_g$ are the rates of loss and gain. These kinds of dissipation can also be found in Ref. \cite{gong2018topologicala,song2019nonhermitiana,mcdonald2022nonequilibrium,yang2022liouvillian}.  Without loss of generality, we assume $t_1,t_2,\gamma_l,\gamma_g$ are all real.

In Eq. (\ref{eq:LME}), terms shaped as $L_{\mu}\rho L_{\mu}^\dagger$ are usually called ``jumping terms''. From a measurement perspective, it represents the continuous
measurements to the system induced by the environment \cite{molmer1993monte}. Here we add a real index $\kappa$ as the quantum jump parameter to characterize the strength that jump occurs. Such structure is named as ``hybrid Liouvillian'' in Ref. \cite{minganti2020hybrid}. In this work, we mainly focus on the two most common cases: the first is the full dynamic of the master equation $(\kappa=1)$, and the second is equivalence to the effective NH dynamic $(\kappa=0)$.
In the literature, the latter case can be achieved by considering a time scale much smaller than the inverse of the dissipation rate ($1/\gamma$) \cite{durrliebliniger}, implementing post-selection \cite{ashida2016quantum} or coherent condensates \cite{gongZenoHallEffect2017}. After the following procedures, the jumping terms in the master equation are often neglected, allowing the system to be effectively described by an equivalent NH Hamiltonian $H_{eff}$. In our model, it takes the form
\begin{equation}
\begin{split}
    H_{eff}=&H-i\sum_m L_m^{\dagger}L_m\\
    =&\sum_n((t_1+\frac{\gamma}{2})c^\dagger_{nA}c_{nB}+(t_1-\frac{\gamma}{2})c^\dagger_{nB}c_{nA}+\\
    &t_2(c^\dagger_{n+1A}c_{nB}+c^\dagger_{nB}c_{n+1A})-\frac{i}{2}\gamma^\prime(c^\dagger_{nA}c_{nA}+c^\dagger_{nB}c_{nB})),\label{eq:Heff}
\end{split}
\end{equation}
here $\gamma=\gamma_l+\gamma_g$, and $\gamma^\prime=\gamma_l-\gamma_g$. Actually, the overall loss $-i\gamma^\prime/2$ is negligible because the total particle numbers of the system $\hat{N}=\sum_n(c^\dagger_{nA}c_{nA}+c^\dagger_{nB}c_{nB})$ can be viewed as a constant  \cite{gong2018topologicala}. Concisely, the schematic diagram of $H_{eff}$ is presented in Fig. \ref{Fig:effmodel}, which is exactly the model studied in Ref. \cite{yao2018edge}. A detailed discussion of its topology will be provided in the Sec. \ref{sec:GBZ}.

\subsection{The third quantization}
It is natural to ask, what may happen if the jumping terms $L_{\mu}\rho L_{\mu}^{\dagger}$ for any $\mu$ can not be neglected. If considering the complete Liouvillian on a doubled Hilbert space \cite{yoshida2020fate}, the dimension of the tensor product space will be the square of the original system $\propto (\text{dim}{H})^2$. 
Given that systems are typically not confined to single-excitation subspaces, it is evidently unacceptable.
In light of these reasons, the third quantization established by Prosen \emph{et al.} may be remarkably convenient in addressing this issue, for it can effectively reduce the dimension to $4N$ \cite{prosen2008third,prosen2010quantization}. Besides, as a general method for dealing with quadratic open systems, it has significant advantages in solving dynamics, steady states, Liouvillian spectra, etc. To start with, we span the free Hamiltonian $H$ and the Lindblad operators $L_\mu$ in terms of the Majorana fermions
\begin{align}
H=\sum_{jk}H_{j,k}w_jw_k,\quad L_\mu=\sum_\mu l_{\mu,m}w_m, 
\end{align}
where 
$w_{2m-1}=c_m+c_m^\dagger, w_{2m}=i(c_m-c_m^\dagger)$, that satisfying anti-commutation relations $\{w_j,w_k\}=2\delta_{j,k}$. The advantage of using $w_j$ is that we can avoid distinguishing between Hermitian conjugates.

The superoperator acts on an operator and generates a new operator. Thus the general method of handling this issue is to transform the operators into vectors and compute within the operator space $\mathcal{K}$ \cite{dzhioev2012nonequilibrium,schmutz1978realtime}. $\mathcal{K}$ is spanned by $4^N$-dimensional canonical basis
$\vert P_{\underline{\alpha}}\rangle$ with
\begin{align}
P_{\underline{\alpha}}=P_{\alpha_1,\alpha_2,\dots,\alpha_{2N}}=w_1^{\alpha_1}w_2^{\alpha_2}\dots w_{2N}^{\alpha_{2N}}.
\end{align}

To seek the explicit form of $\hat{\mathcal{L}}$, one can introduce a set of adjoint annihilation and creation Fermi maps $\hat{c}^\dagger_j,\hat{c}_j$ (named as ``a-fermions'', we use a ``hat'' for distinction), defined as 
\begin{equation}
    \hat{c}_j\vert P_{\underline{\alpha}}\rangle=\delta_{\alpha_j,1}\vert w_jP_{\underline{\alpha}}\rangle,\quad\hat{c}^\dagger_j\vert P_{\underline{\alpha}}\rangle=\delta_{\alpha_j,0}\vert w_jP_{\underline{\alpha}}\rangle\quad\alpha_j\in\{0,1\},\label{eq:anni&cre}
\end{equation}
which obey the canonical anti-commutation relations $\{\hat{c}_j,\hat{c}^\dagger_k\}=\delta_{jk}$.

With the adjoint Fermi maps $\hat{c}_j,\hat{c}^\dagger_j$, each part of the Lindblad maps can be written out \cite{prosen2008third}.  First, the a-Fermi map of unitary part of Liouvillian $\mathcal{L}_0\rho:=-i\left[H,\rho\right]$ writes
\begin{align}
\hat{\mathcal{L}}_0=&-4i\sum_{j,k}\hat{c}_j^\dagger H_{j,k}\hat{c}_k,\label{eq:L0}
\end{align}
the second part is the dissipation part $\mathcal{L}_{diss}\rho:=\mathcal{L}_{eff}+\kappa\mathcal{L}_{jump}=\sum_\mu(2\kappa L_\mu\rho L^\dagger_\mu-\rho L^\dagger_\mu L_\mu-L^\dagger_\mu L_\mu\rho)$, here we divided it into $\mathcal{L}_{eff}$ and $\mathcal{L}_{jump}$, induced by the effective theory and jumps, respectively. In the even parity subspace
\begin{align}
&\hat{\mathcal{L}}_{eff}=\sum_{j,k,\mu}2l_{\mu,j}l_{\mu,k}^*(\hat{c}_j^\dagger\hat{c}_k^\dagger+\hat{c}_j\hat{c}_k),\label{eq:Leff}\\
    &\hat{\mathcal{L}}_{jump}=\sum_{j,k,\mu}2l_{\mu,j}l_{\mu,k}^*(\hat{c}^\dagger_j\hat{c}^\dagger_k-\hat{c}^\dagger_j\hat{c}_k-\hat{c}^\dagger_k\hat{c}_j-\hat{c}_j\hat{c}_k).\label{eq:Ljump}
\end{align}

On the contrary, when considering the full dynamic ($\kappa=1$), $\hat{\mathcal{L}}_{diss}$ will become
\begin{equation}
\hat{\mathcal{L}}_{diss}=\sum_{j,k,\mu}l_{\mu,j}l_{\mu,k}^*(4\hat{c}^\dagger_j\hat{c}^\dagger_k-2\hat{c}^\dagger_j\hat{c}_k-2\hat{c}^\dagger_k\hat{c}_j). \label{eq:even}
\end{equation}

After reducing to normal master modes (NMMs) $\hat{a}_{2j-1}=(\hat{c}_{j}+\hat{c}_j^\dagger)/\sqrt{2},\hat{a}_{2j}=i(\hat{c}_{j}-\hat{c}_j^\dagger)/\sqrt{2}$ \cite{prosen2008third}, the Liouvillian becomes bilinear
\begin{equation}
\hat{\mathcal{L}}=\sum_{j,k}^{4N}\textbf{A}_{j,k}\hat{a}_j\hat{a}_k,
\end{equation}
the $\textbf{A}$ is usually named as \emph{shape matrix}, with entries
\begin{align}
    &\textbf{A}_{2j-1,2k-1}=-2i{H}_{jk}-\sum_{\mu}l_{\mu,j}l_{\mu,k}^*+\sum_{\mu}l_{\mu,k}l_{\mu,j}^*,\nonumber\\
    &\textbf{A}_{2j-1,2k}=-2{H}_{jk}+2i\sum_{\mu}l_{\mu,k}l_{\mu,j}^*,\nonumber\\
    &\textbf{A}_{2j,2k-1}=2{H}_{jk}-2i\sum_{\mu}l_{\mu,j}l_{\mu,k}^*,\nonumber\\
    &\textbf{A}_{2j,2k}=-2i{H}_{jk}+\sum_{\mu}l_{\mu,j}l_{\mu,k}^*-\sum_{\mu}l_{\mu,k}l_{\mu,j}^*,
    \end{align}
it is not hard to show that $\textbf{A}$ is antisymmetric
    $\textbf{A}=-\textbf{A}^T$.

\subsection{The \emph{shape matrix} of the model}
From the procedures in the previous subsection, we can now turn to the shape matrix $\textbf{A}$ of our model, which is constituted of two parts: (i)
$\textbf{A}_{0}$ that is derived from the free Hamiltonian $H$. (ii) $\textbf{A}_{diss}$ that describes the dissipation part
\begin{align}
\textbf{A}_{0}=\frac{1}{2}\left(
\begin{matrix}
& t_1T & & & & t_2T\\
t_1T & & t_2T & & &\\
& t_2T & & t_1T & &\\
& &  \ddots & & \ddots &\\
t_2T & & & &  & \\
\end{matrix}
\right),\label{eq:A0}
\end{align}
here $T=-i\sigma_y\otimes I$, where $\sigma_y$ is the Pauli matrix and $I$ is the identity operator. For simplicity, here we consider the case of pure loss $(\gamma_g=0)$ (the case that contains both loss and gain will be discussed afterwards)
\begin{align}
    \textbf{A}_{diss}=\frac{1}{2}\left(\begin{matrix}
    A & B & & & &\\
    -B & A & & & &\\
    & & A & B & & \\
    & & & \ddots & \ddots &\\
    & & & & -B & A
    \end{matrix}\right),
    \label{eq:Adiss}
\end{align}
the matrices $A$ and $B$ write
 \begin{align}
&A=-\frac{\gamma_l}{2}\left(\begin{matrix}
0 & -i\kappa & -i & \kappa\\
i\kappa & 0 & \kappa & i\\
i & -\kappa & 0 & -i\kappa\\
-\kappa & -i & i\kappa & 0\\
\end{matrix}\right),
B=-\frac{\gamma_l}{2}\left(\begin{matrix}
i & -\kappa & 0 & -i\kappa\\
-\kappa & -i & i\kappa & 0\\
0 & i\kappa & i & -\kappa\\
-i\kappa & 0 & -\kappa & -i\\
\end{matrix}\right).
\end{align}

Suppose the model is under period boundary condition (PBC), it is relatively simple to transform into the momentum space. By applying the Fourier transformation, the Bloch shape matrix $\textbf{A}(k)$ reads
\begin{align}
    \textbf{A}(k)=\frac{1}{2}\left(\begin{matrix}
    A & (t_1+t_2e^{-ik})T+B\\
    (t_1+t_2e^{ik})T-B & A
    \end{matrix}\right).\label{eq:Ak}
\end{align}

On the other hand, when in the absence of jumps ($\kappa=0$), the Bloch form
shape matrix will be the same as Eq. (\ref{eq:Ak}), except that $A,B$ are substituted by $A^\prime,B^\prime$, respectively
\begin{align}
 \textbf{A}^\prime(k)=\frac{1}{2}\left(\begin{matrix}
 A^\prime & (t_1+t_2e^{-ik})T+B^\prime\\
 (t_1+t_2e^{ik})T-B^\prime & A^\prime
 \end{matrix}\right),\label{eq:Aprime}
\end{align}
where
\begin{align}
A^\prime=-\frac{\gamma_l}{2}\left(\begin{matrix}
0 & 0 & -i & 0\\
0 & 0 & 0 & i\\
i & 0 & 0 & 0\\
0 & -i & 0 & 0
\end{matrix}\right),
B^\prime=-\frac{\gamma_l}{2}\left(\begin{matrix}
i & 0 & 0 & 0\\
0 & -i & 0 & 0\\
0 & 0 & i & 0\\
0 & 0 & 0 & -i
\end{matrix}\right).
\end{align}

It's worth noting that, the Liouvillian $\mathcal{L}$ will remain invariant under a transformation $H \rightarrow H + cI$, where $I$ denotes the identity operator, and $c$ is a complex number \cite{zloshchastiev2014comparisona}. Consequently, in this context, the term $A^\prime$ can be neglected, as it represents the overall loss, which can be considered as an unimportant energy shift.

\section{\label{sec:symmetry}Symmetry} 
Symmetry representation has always played a crucial role in the evolution of band theory. Starting with the model of noninteracting fermions, its symmetry could be sorted into one of Altland-Zirnbauer (AZ) symmetry class \cite{altland1997nonstandard,kitaev2009periodic,ryu2010topologicala}, which based on the presence or absence of three fundamental symmetries: time-reversal symmetry (TRS), particle-hole symmetry (PHS), and chiral symmetry (or sublattice symmetry). The various combinations of these symmetries will result in a ten-fold classification
\begin{subequations}
\begin{align}
&\text{TRS:}H=U_TH^*U_T^\dagger, U_TU_T^*=\pm I,\label{eq:TRS}\\
&\text{PHS:}H=-U_CH^*U_C^\dagger, U_CU_C^*=\pm I,\label{eq:PHS}\\
&\text{chiral:}H=-U_SHU_S^\dagger, U_S^2=I,\label{eq:chiral}
\end{align}
\end{subequations}
here $U_T,U_C,U_S$ are unitary matrices of TRS, PHS and chiral symmetry. 

When moving to NH systems, the introduction of non-Hermiticity will significantly broaden the classification, because the equivalence between lattice symmetry and chiral symmetry in Hermitian systems may break down. It has been demonstrated that there are a total of 38 distinct symmetry classes in NH systems \cite{kawabataSymmetryTopologyNonHermitian2019}. However, even in such cases, when the Lindblad operator exhibits linearity with respect to fermions, similar tenfold symmetry classification will still perserve, owing to the physical constraints imposed by the Lindbladian spectrum
\cite{lieu2020tenfold,he2022topological}. 

Precisely, based on the above considerations, by using Pauli matrices, we can rewrite Eq. (\ref{eq:Ak}) as
\begin{align}
&\textbf{A}(k)=(-i(t_1+t_2\cos{k})\sigma_x\otimes(\sigma_y\otimes I_2)-
it_2\sin{k}\sigma_y\otimes\nonumber\\
&(\sigma_y\otimes I_2)-\frac{i\gamma}{2}\sigma_y\otimes(I_2\otimes(i\sigma_z-\kappa\sigma_x)+i\kappa\sigma_y\otimes\sigma_y)-\frac{\gamma}{2}\kappa I_2\nonumber\\
&\otimes I_2\otimes\sigma_y+\frac{i\gamma}{2}I_2\otimes\sigma_y\otimes(i\sigma_z-\kappa\sigma_x))/2.\label{eq:pauliAk}
\end{align}
Obviously, it is easy to prove that Eq. (\ref{eq:pauliAk}) satisfies TRS in Eq. (\ref{eq:PHS}), because when $U_T=I_2\otimes\sigma_y\otimes\sigma_x$, we have
 \begin{align}
     U_T\textbf{A}^*(k)U_T^{-1}=\textbf{A}(-k),
 \end{align}
and $U_TU^*_T=-I$.

Conversely, when there is no jump and assuming the overall loss is neglected, Eq. (\ref{eq:Aprime}) becomes
\begin{align}
\textbf{A}^\prime(k)=&(-i(t_1+t_2\cos{k})\sigma_x\otimes(\sigma_y\otimes\nonumber I_2)-it_2\sin{k}\sigma_y\otimes\\
&(\sigma_y\otimes I_2)+\frac{\gamma}{2}\sigma_y\otimes I_2\otimes\sigma_z)/2,\label{eq:pauliAkprime}
\end{align}
which satisfies chiral symmetry in Eq. (\ref{eq:chiral}) by selecting $S=\sigma_z\otimes I_2\otimes I_2$
 \begin{align}
  \textbf{A}^\prime(k)=U_S\textbf{A}^\prime(k)U_S^{-1}.
  \end{align}
 Meanwhile, $\textbf{A}^\prime(k)$ also has other symmetries (TRS in Eq. (\ref{eq:TRS}) and PHS in Eq. (\ref{eq:PHS})), because
 \begin{align}
   \textbf{A}^\prime(k)=U_T\textbf{A}^\prime(-k)U_T^{-1}, \\
   \textbf{A}^\prime(k)=-U_C\textbf{A}^\prime(-k)U_C^{-1},
 \end{align}
 where $U_T=I_2\otimes\sigma_y\otimes\sigma_x$ and $U_C=\sigma_z\otimes \sigma_y\otimes\sigma_x$. 
 It is evident that these two cases are protected by distinct symmetries, because the introduction of quantum jumping terms will break the chiral symmetry. A comprehensive analysis of their properties will be proceeded in the subsequent sections.

\begin{figure}[b]
\centering
\includegraphics[width=0.8\textwidth]{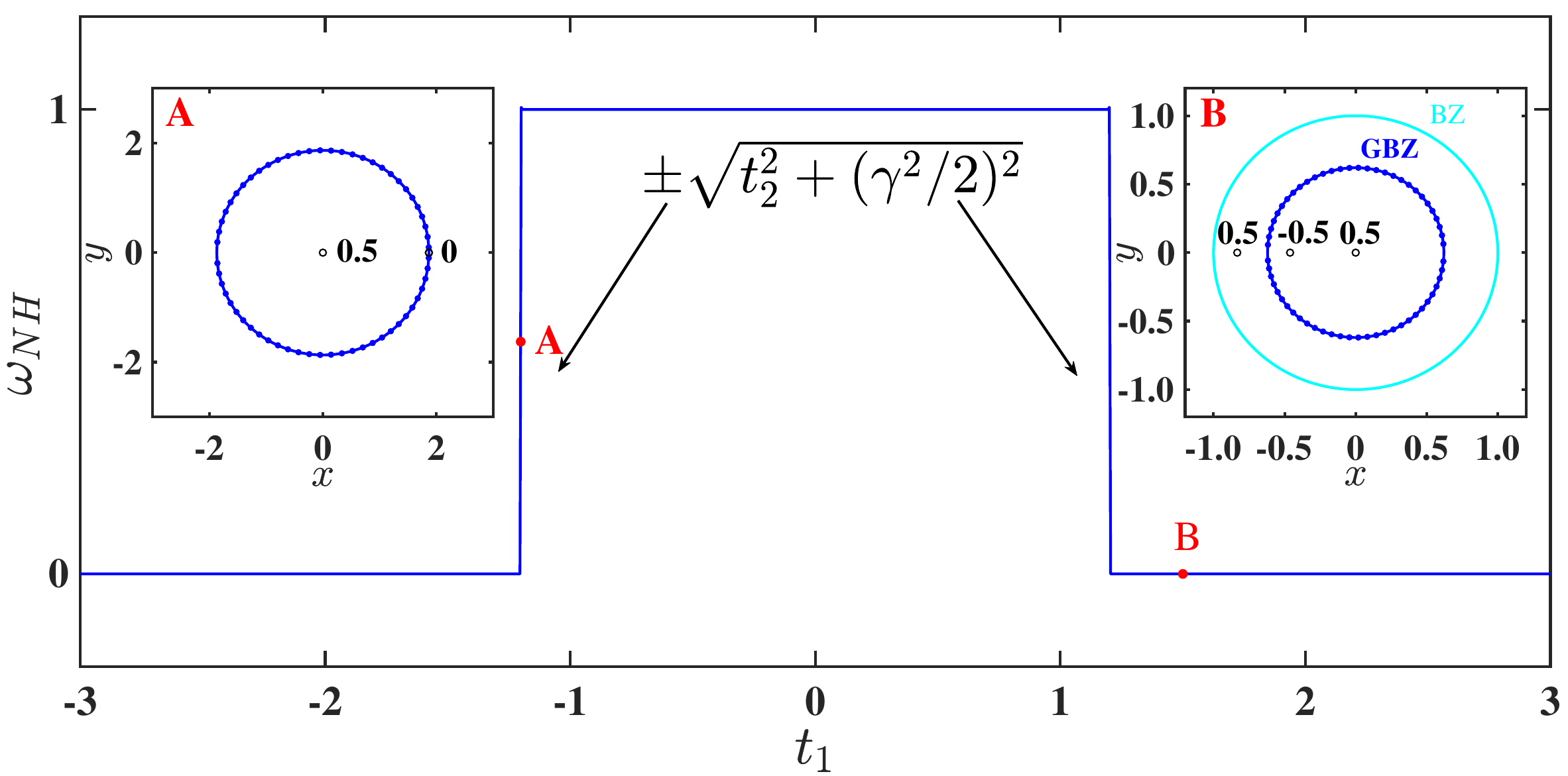}
    \caption{The non-Bloch topological invariant $\omega_{NH}$ of the effective model in Fig. \ref{Fig:effmodel} and its corresponding GBZ (blue dots and circles in insets). The transition points are at
    $t_1=\pm\sqrt{(\gamma/2)^2+t_2^2}$, which cannot be predicted by the conventional BZ (cyan circles as in insect B).
    In inset A, $t_1=-\sqrt{(\gamma/2)^2+t_2^2}\approx -1.2019$, while in B, $t_1=1.5$. As a comparison, the scattered points are the numerical calculation ($L=30$) of Eq. (\ref{eq:HeffE}), coincides with the analytical region calculated by aGBZ. The poles and their corresponding residues according to Eq. (\ref{eq:h12}) are also shown. The parameters are chosen as $t_2=1,\gamma=4/3$.}\label{Fig:wNH}
\end{figure}  

\section{Generalized
Brillouin Zone \label{sec:GBZ}}

In this section, we begin with the re-establish of the bulk-edge correspondence (BBC) in NH systems and introduce the theory of the generalized Brillouin zone (GBZ).
In Sec. \ref{sec:GBZ_NH}, by utilizing the concept of the auxiliary generalized Brillouin zone (aGBZ), we recall the analysis of GBZ associated with the effective NH model (Fig. \ref{Fig:effmodel}). To characterize the effect of quantum jumps, we reexamined this issue from the open description (Fig. \ref{Fig:model}) in Sec. \ref{sec:aGBZ_nojump} and \ref{sec:GBZ_jump}, when there lacks or experiences quantum jumps, respectively.

In Hermitian topological systems, the concept of invariants is intricately connected to the Brillouin zone (BZ). Within this context, the most fundamental property is the BBC, which establishes a connection between the edge states and bulk topological invariant. Precisely, the
existence of boundary states under open boundary conditions (OBC) can be faithfully predicted by the bulk topological invariants. However, the restoration of this correspondence in NH systems has encountered difficulties. Even for the simplest SSH models, are controversially debated and discussed \cite{wang2020defective,yin2018geometrical,xiong2017why,ashida2020nonhermitiana}. Initially, it was believed that NH systems would lead to a complete breakdown of BBC. To restore this relationship, substantial efforts were made, including introducing fractional winding numbers \cite{lee2016anomalous}, multiple winding numbers \cite{yin2018geometrical}, and the biorthogonal polarization approach \cite{kunst2018biorthogonal}. However, the topological winding number defined in these manners often encounter difficulties in predicting the appearance of the phase transition points
\cite{yao2018edge,xiong2017why}.

Fortunately, with the establishment of the generalized Brillouin zone theory, this issue has been partially resolved to some extent \cite{yao2018edge,yokomizo2019nonBloch}.
As it fundamentally reveals the ``skin effect'' \cite{yao2018nonhermitian,okumaTopologicalOriginNonHermitian2020a}, where the wave functions of NH systems may concentrate near the boundaries, that plays a crucial role in establishing the ``non-Bloch bulk-boundary correspondence''. 
Furthermore, due to their sensitivity to boundary conditions, the conventional BZ may breakdown. Instead, the wave vector $k$ may become a complex number, with the trajectory of $e^{ik}$ in the complex plane forming the so-called ``GBZ''. Generally, the GBZ is no longer a unit circle, and the BZ can be seen as a special case in the Hermitian limit. Despite a diversity of viewpoints persists, the theory of GBZ sparked a wave of research on topological non-Hermitian systems \cite{kawabata2020nonBloch,wu2022connectionsa,zhang2020correspondence,guo2021nonhermitian}, and pave the way for further investigations.

\subsection{The GBZ of NH Hamiltonian}\label{sec:GBZ_NH}
 As a starting point, we begin with the effective model in Fig. \ref{Fig:effmodel}. Considering pure loss condition $\gamma_g=0, \gamma_l=\gamma$, the system is identical to the asymmetric NH model discussed in Ref. \cite{yao2018edge}, with the following conclusion holds
 \begin{align}
     \left[(t_1-\frac{\gamma}{2})+t_2\beta\right]\left[(t_1+\frac{\gamma}{2})+t_2\beta^{-1}\right]=E^2,\label{eq:HeffE}
 \end{align}
 here $E$ is the bulk-band  energy under the OBC, and $\beta$ denotes the replaced Bloch phase factor that uses the replacement $\beta\rightarrow e^{ik}, 1/\beta\rightarrow e^{-ik}$.
 
 To solve for $\beta$, the OBC spectrum $\{E\}$ shall be calculated first through numerical diagonalization. For each $E$, a set of $2N$ equations in terms of $\beta$ will be established. Under the requirement of the continuous band condition, a pair of solutions with the same modulus will constitute the GBZ of the system
 (for the procedures of calculating GBZ, see \ref{app:gbz} for more details). However, such a method is highly sensitive to the model size and calculation precision. Inadequate precision can lead to numerical diagonalization errors, while excessive precision will result in unacceptable computation time. Furthermore, it is also challenging to determine whether there exists self-intersection points on the GBZ.
 
 To overcome the aforementioned limitations, we use the \emph{auxiliary Generalized Brillouin Zone} (aGBZ) theory presented in Ref. \cite{yang2020nonhermitiana}, which employs the concept of \emph{resultant} to eliminate the unwanted parameters from the equations. For instance, in this context,
 we exclude the variables $E$ and the phase angle $\theta$ from the equation. By exploiting the feature of resultant, we could derive the analytical expression in terms of $\beta$ (for further details, refer to \ref{app:aGBZ})
 \begin{align}
     |\frac{\gamma}{2}+t_1|(x^2+y^2)-|\frac{\gamma}{2}-t_1|=0,\label{eq:gbz}
 \end{align}
 here $x=\text{Re}\,\beta$ and $y=\text{Im}\,\beta$, whose trajectories in the complex plane form
 the GBZ. In the insets of Fig. \ref{Fig:wNH}, we display the GBZ of the effective model, with the methods of aGBZ (blue solid line) and numerical results
(scattered dots, by Eq. (\ref{eq:HeffE})). Their results are consistent, except that numerical results will become extremely loose near the real axis. In contrast, the analytical solution provided by the aGBZ method has more advantages for subsequent calculations. Such aspects will become increasingly evident when dealing with more intricate systems.

\subsection{The aGBZ of shape matrix when no jump}\label{sec:aGBZ_nojump}
 
The topological formulation of open system often serves as an extension of NH systems, with focusing on its non-equilibrium steady states (NESS) or transient dynamics, has become a valuable topic of current research
\cite{dangel2018topological,lieu2020tenfold,yoshida2020fate,rivas2013density}. In addition to the nature that both the system's Liouvillian \cite{minganti2019quantum} and shape matrix exhibit non-Hermiticity, the localizability of the NMMs have also been observed as in the NH systems \cite{vancaspel2019dynamical}. With these similarities, in the same fashion, we can analogously write the eigenequation $\textbf{A}\vert \psi\rangle=E\vert \psi\rangle$ as
\setcounter{MaxMatrixCols}{16}
\begin{equation}
\resizebox{0.95\hsize}{!}{$\begin{aligned}
    \left(\begin{matrix}
    \frac{i\gamma}{2} & -\frac{\gamma}{2}\kappa & -t_1 & -\frac{i\gamma}{2}\kappa & 0 & \frac{i\gamma}{2}\kappa & \frac{i\gamma}{2} & -\frac{\gamma}{2}\kappa & 0 & 0 & -t_2 & 0 & 0 & 0 & 0 &0\\
    -\frac{\gamma}{2}\kappa & -\frac{i\gamma}{2} & \frac{i\gamma}{2}\kappa & -t_1 & -\frac{i\gamma}{2}\kappa & 0 & -\frac{\gamma}{2}\kappa & -\frac{i\gamma}{2} & 0 & 0 & 0 & -t_2 & 0 & 0 & 0 & 0\\
    t_1 & \frac{i\gamma}{2}\kappa & \frac{i\gamma}{2} & -\frac{\gamma}{2}\kappa & -\frac{i\gamma}{2} & \frac{\gamma}{2}\kappa & 0 & \frac{i\gamma}{2}\kappa & t_2 & 0 & 0 & 0 & 0 & 0 & 0 & 0\\
    -\frac{i\gamma}{2}\kappa & t_1 & -\frac{\gamma}{2}\kappa & -\frac{i\gamma}{2} & \frac{\gamma}{2}\kappa & \frac{i\gamma}{2} & -\frac{i\gamma}{2}\kappa & 0 & 0 & t_2 & 0 & 0 & 0 & 0 & 0 & 0\\
    0 & 0 & 0 & 0 & 0 & 0 & -t_2 & 0 & 0 & \frac{i\gamma}{2}\kappa & \frac{i\gamma}{2} & -\frac{\gamma}{2}\kappa & -\frac{i\gamma}{2} & \frac{\gamma}{2}\kappa & -t_1 & \frac{i\gamma}{2}\kappa\\
    0 & 0 & 0 & 0 & 0 & 0 & 0 & -t_2 & -\frac{i\gamma}{2}\kappa & 0 & -\frac{\gamma}{2}\kappa & -\frac{i\gamma}{2} & \frac{\gamma}{2}\kappa & \frac{i\gamma}{2} & -\frac{i\gamma}{2}\kappa & -t_1\\
    0 & 0 & 0 & 0 & t_2 & 0 & 0 & 0 & -\frac{i\gamma}{2} & \frac{\gamma}{2}\kappa & 0 & \frac{i\gamma}{2}\kappa & t_1 & -\frac{i\gamma}{2}\kappa & -\frac{i\gamma}{2} & \frac{\gamma}{2}\kappa\\
    0 & 0 & 0 & 0 & 0 & t_2 & 0 & 0 & \frac{\gamma}{2}\kappa & \frac{i\gamma}{2} & -\frac{i\gamma}{2}\kappa & 0 & \frac{i\gamma}{2}\kappa & t_1 & \frac{\gamma}{2}\kappa & \frac{i\gamma}{2}
\end{matrix}\right)\left(\begin{matrix}
\psi_{nA1}\\\psi_{nA2}\\\psi_{nA3}\\ \psi_{nA4}\\\psi_{nB1}\\\psi_{nB2}\\ \psi_{nB3}\\\psi_{nB4}\\\psi_{n+1A1}\\\psi_{n+1A2}\\\psi_{n+1A3}\\ \psi_{n+1A4}\\\psi_{n+1B1}\\\psi_{n+1B2}\\ \psi_{n+1B3}\\\psi_{n+1B4}\end{matrix}\right)=
    E\left(\begin{matrix}
\psi_{nA1}\\\psi_{nA2}\\\psi_{nA3}\\ \psi_{nA4}\\\psi_{nB1}\\\psi_{nB2}\\ \psi_{nB3}\\\psi_{nB4}\end{matrix}\right)
\end{aligned}$},\label{eq:matrixjump}
\end{equation}
here $\vert \psi\rangle=(\psi_{1A1},\psi_{1A2},\psi_{1A3},\psi_{1A4},\psi_{1B1},\psi_{1B2},\psi_{1B3},\psi_{1B4},\dots)^T$ is the
right eigenvector of the shape matrix within the basis of NMMs. $\psi_{m\sigma n}$ is the wave function at sites $\sigma=A,B$, $m$ is the index of the cell, and
$n=1,2,3,4$ stands for the NMMs \footnote{For a quadratic system of $n$ fermions, its adjoint Hermitian Majorana maps $\hat{a}_r=\hat{a}_r^\dagger$ takes $r=1,2,\dots,4n$.}. 

\begin{figure}[htbp]
\centering
\subfigure{
\label{Fig:multi1}
\includegraphics[width=0.36\textwidth]{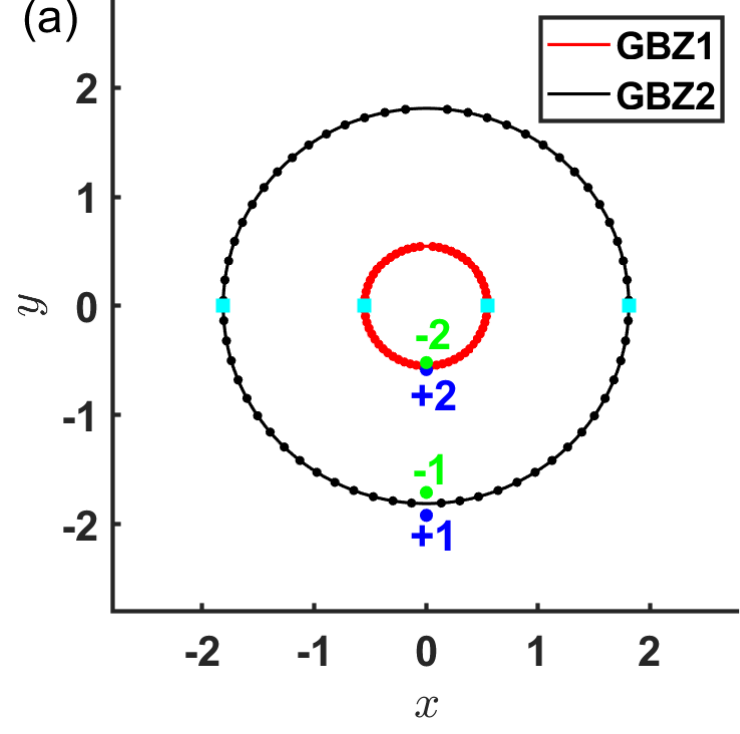}}\subfigure{
\label{Fig:multi2}
\includegraphics[width=0.36\textwidth]{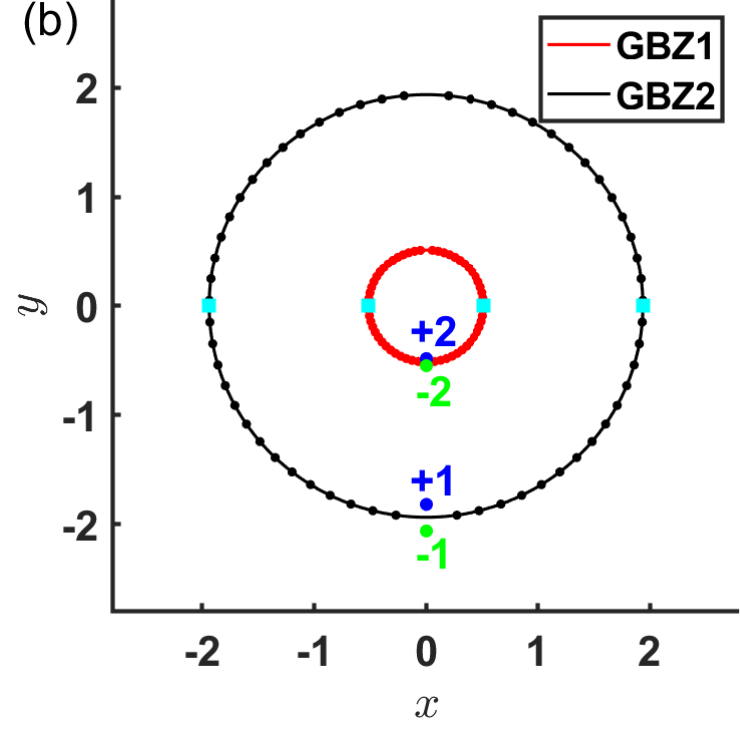}}
\subfigure{
\label{Fig:spectraA}
\centering
\includegraphics[width=0.36\textwidth]{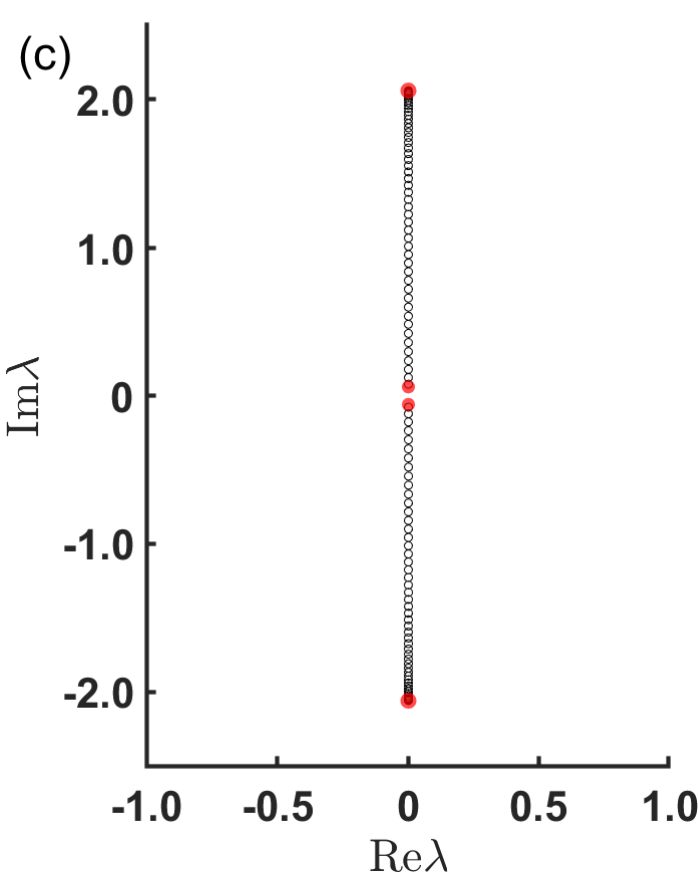}}\subfigure{
\label{Fig:spectraB}
\includegraphics[width=0.36\textwidth]{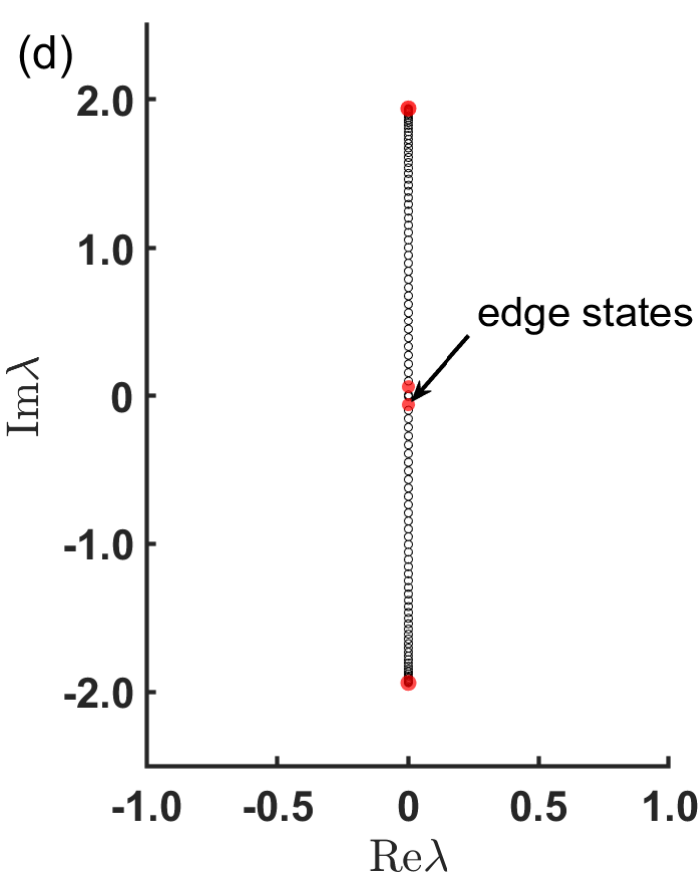}}
    \caption{(a-b) The GBZ of jumping absent model $\textbf{A}^\prime$. The red and black solid lines represent the arcs of GBZ1 and GBZ2, respectively, while the scattered dots denote numerical results obtained by solving Eq.  (\ref{eq:analysisGBZ}) with $L=30$.
    The cyan squares represent the self-conjugate points ($\beta_M=\beta_{M+1}$), whereas the blue and green points denote the zeros of $\text{det}(h_1)$ and $\text{det}(h_2)$. (c-d) The spectrum of $\textbf{A}^\prime$ around transition point (black cycles, $L=50$). In (a,c) $t_1=\sqrt{(\gamma/2)^2+t^2_2}+0.05\approx1.2519$, and in (b,d) $t_1=\sqrt{(\gamma/2)^2+t^2_2}-0.05\approx1.1519$. $t_2=1,\gamma=4/3$. The red dots represent the self-conjugate points, which restricts the range of the energy spectrum under OBC. It is clear that there shows zero eigen-energy in (d) (eightfold degenerate).}
\end{figure}
 
Starting with the jumping absence case ($\kappa=0$), the system does not exhibit any quantum jump events and is dynamically equivalent to the effective Hamiltonian $H_{eff}$. Due to spatial periodicity Eq. (\ref{eq:sumpsi}), we can obtain a set of homogeneous linear equations. In order to have non-zero solutions, the determinant of the coefficient matrix shall take zero, which yields
\begin{subequations}
\begin{align}
    E^2-\gamma^2/4+t_1^2+\frac{\gamma t_2}{2\beta}-\frac{\beta\gamma t_2}{2}+\frac{t_1t_2}{\beta}+\beta{t_1t_2}+t_2^2=0,\label{eq:analysisGBZ1}\\ E^2-\gamma^2/4+t_1^2-\frac{\gamma t_2}{2\beta}+\frac{\beta\gamma t_2}{2}+\frac{t_1t_2}{\beta}+\beta{t_1t_2}+t_2^2=0.\label{eq:analysisGBZ2}
\end{align}\label{eq:analysisGBZ}
\end{subequations}

Actually, the two equations are interrelated via a gauge $\gamma\rightarrow-\gamma, \beta\rightarrow1/\beta$, signifying the propagation of the Bloch wave function in opposite directions at each ends. By eliminating the addition degrees $E,\theta$, the irreducible factorizations of $R_t\left[\text{Re}(G),\text{Im}(G)\right]=0$ present the trajectory of aGBZ. For Eq. (\ref{eq:analysisGBZ1}), it indicates

\begin{align}
    |\gamma/2-t_1|(x^2+y^2)-|\gamma/2+t_1|=0,\label{eq:aGBZ1}
 \end{align}
    while Eq. (\ref{eq:analysisGBZ2}), we have
\begin{align}
    |\gamma/2+t_1|(x^2+y^2)-|\gamma/2-t_1|=0,\label{eq:aGBZ2}
\end{align}
same as previous, $x,y$ take $x=\text{Re}\,\beta$ and $y=\text{Im}\,\beta$.

For a given set of parameters, Eq. (\ref{eq:aGBZ1})-(\ref{eq:aGBZ2}) actually correspond to distinct aGBZ regions. It is because for multiband systems, the projection of different branches of the spectrum onto the complex plane may correspond to distinct sub-GBZs, thereby forming multiple GBZs. However,
we must emphasize a verification is required in order
to avoid incremental roots that may be introduced during the elimination (see \ref{app:multigbz} for the definition of self-conjugate points in aGBZ, as well as the concept of the end points to delineate the actual areas of GBZ).

\begin{figure}[htbp]
\centering
\includegraphics[width=0.8\textwidth]{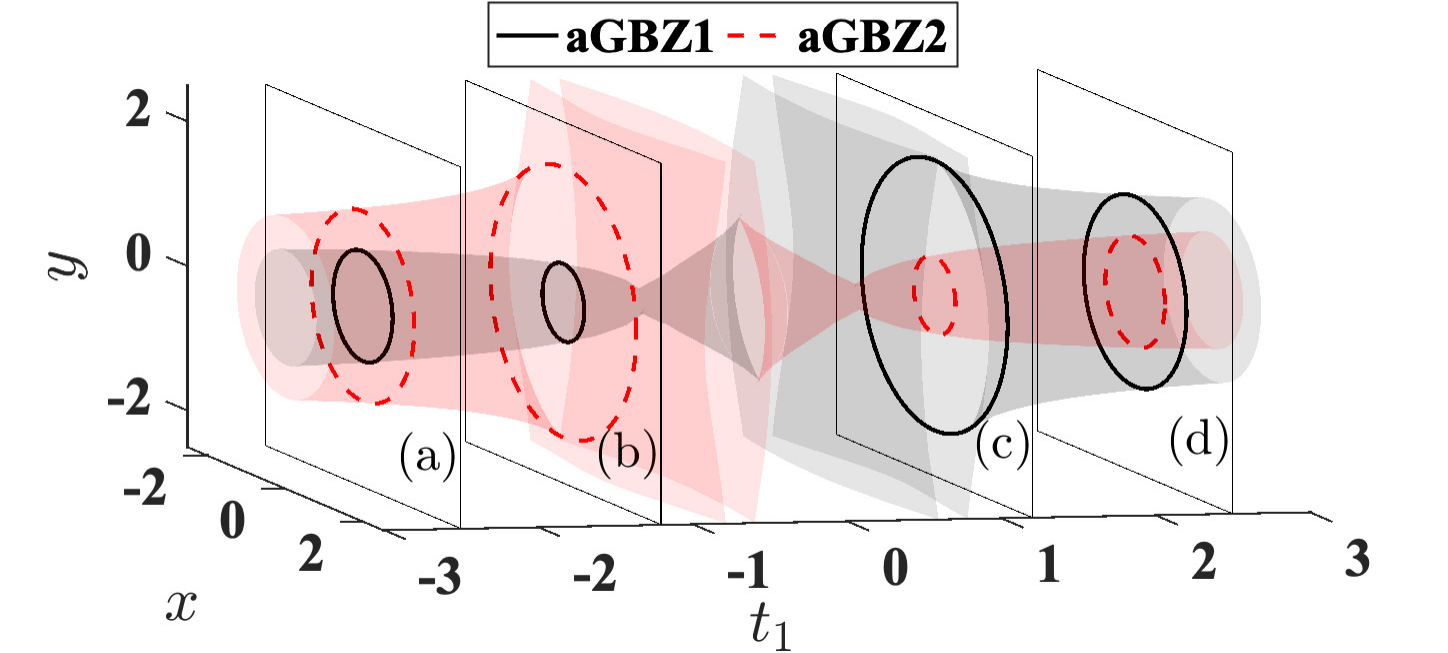}
    \caption{The aGBZ of open quantum description (Eq. (\ref{eq:Aprime})) without jump. The different colors represent arcs of aGBZ bands in Eq. (\ref{eq:aGBZ1}), (\ref{eq:aGBZ2}). Parameters are chosen as (a) $t_1=-2.5$, (b) $t_1=-\sqrt{(\gamma/2)^2+t_2^2}$, (c) $t_1=\sqrt{(\gamma/2)^2+t_2^2}$, (d) $t_1=2.5$, and $t_2=1,\gamma=4/3$.}\label{Fig:GBZ12}
\end{figure}

By utilizing the technique introduced in \ref{app:multigbz}, in Fig. \ref{Fig:multi1}-\ref{Fig:spectraB}, we show the end points $E_0$ of the spectrum (red points) by solving Eq. (\ref{eq:cpoints}) around the transition points when $t_1=\sqrt{(\gamma/2)^2+t_2^2}\pm 0.05$. The corresponding self-conjugate points $\beta_0$ are marked in Fig. \ref{Fig:multi1},\ref{Fig:multi2} with cyan squares. All the self-conjugate points $\beta_0$ that appear on the trajectory of the aGBZ must form GBZ \cite{yang2020nonhermitiana}. Thus both branches of the closed loop are either aGBZ or GBZ.
The corresponding schematic of aGBZ that changes as a function of $t_1$ is illustrated in
Fig. \ref{Fig:GBZ12}.

\subsection{The GBZ of shape matrix when jump presented}\label{sec:GBZ_jump}

To visualize the effect of jumping terms, we reconsider the eigen-function Eq. (\ref{eq:matrixjump}) ($\kappa\ne0$). Similarly, we could establish the relationship between $E$ and $\beta$
\begin{subequations}
\begin{align}
    E^2-\gamma E+\frac{\gamma t_2}{2\beta}-\frac{\gamma t_2}{2}\beta+\frac{t_1t_2}{\beta}+\beta t_1t_2+t_1^2+t^2_2=0,\label{eq:Djump1}\\
    E^2+\gamma E-\frac{\gamma t_2}{2\beta}+\frac{\gamma t_2}{2}\beta+\frac{t_1t_2}{\beta}+\beta t_1t_2+t_1^2+t^2_2=0
,\label{eq:Djump2}
\end{align}\label{eq:Djump}
\end{subequations}
which differs from Eq. (\ref{eq:analysisGBZ}) only by a constant term $\pm\gamma/2$, originating from the overall loss of the dynamic (as in Eq. (\ref{eq:Heff})). It is worth noting that this equation is independent of $\kappa$, which implies that the GBZ remains unchanged under different jumping strengths.
In this regard, we will discuss and elaborate on this matter in Sec. \ref{sec:winding number}.

\section{Topological invariants\label{sec:winding number}}

In this section, we will unravel the topology of the system. In Sec. \ref{sec:topo_NH}, \ref{sec:topo_withoutjump}, from the perspective of  
NH and the open system description, we
calculate their corresponding topological invariants and extend our analysis to the full dissipative system in Sec. \ref{sec:topo_withjump}. In Sec. \ref{sec:dynamical_signatures}, an explanation based on transient dynamics is presented. Finally, we validate BBC in Sec. \ref{sec:BBC}.

\begin{figure}[b]
\centering
\includegraphics[width=0.8\textwidth]{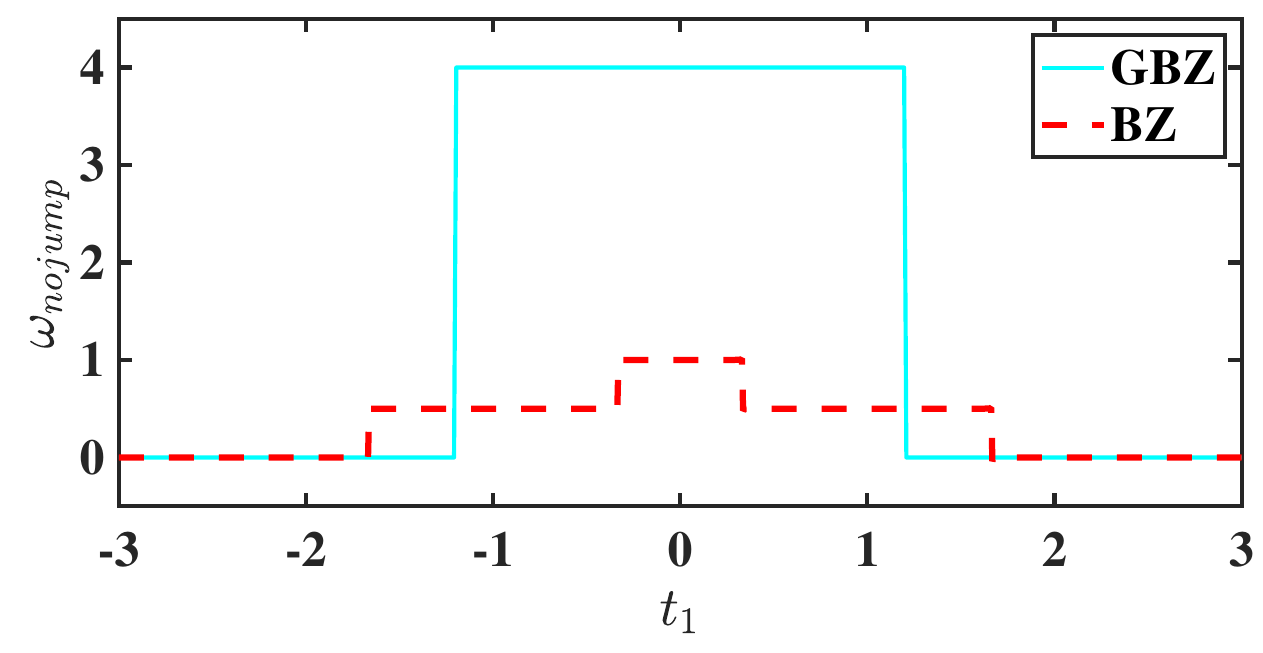}
    \caption{The winding number (cyan solid line) calculated via GBZ (Eq. (\ref{eq:omeganojump})), when there is no quantum jump. For comparison purposes, results obtained by conventional BZ are also presented (red dashed line, Eq. \ref{eq:omega_njBZ}).}\label{Fig:windingnojump}
\end{figure}

\subsection{The NH Hamiltonian}\label{sec:topo_NH}

From Ref. \cite{yao2018edge}, the Bloch form Hamiltonian $H_{eff}$ is topologically protected by chiral symmetry $\sigma_zH_{eff}(k)\sigma_z=-H_{eff}(k)$,
whose winding number can be defined by the ``Q'' matrix method \cite{chiu2016classification}: by choosing a proper basis that makes the symmetry operator diagonal, the Hamiltonian $H$ can be written as
\begin{align}
    H=\left(\begin{matrix}
      & h_+\\
    h_- &
    \end{matrix}\right),\label{eq:Hh12}
\end{align}
here $h_{\pm}$ is typically a matrix. For Hermitian cases $h_-=h_+^*$, while for NH systems, the relationship between $h_+$ and $h_-$ may be determined by the special internal symmetry of the system \cite{gong2018topologicala}. Based on the intrinsic properties of matrix $H$, the following relationship holds \cite{yokomizo2019nonBloch}
\begin{align}
    h_+h_-\vert \psi^R_\mu\rangle=E^2_\mu(\beta)\vert \psi^R_\mu\rangle,
\end{align}
where the $\{\vert\psi^R_\mu\rangle\},\{\vert\psi^L_\mu\rangle\}$ is the set of right and left eigenvectors and $\langle\psi^L_\nu\vert\psi^R_\mu\rangle=\delta_{\mu,\nu}$. Hence the $Q$ matrix can be calculated
\begin{align}
Q=\left(
    \begin{matrix}
      & q(\beta)\\
    q^{-1}(\beta) &
    \end{matrix}
    \right),
\end{align}
where $q(\beta)=\sum_\mu q_\mu(\beta)=\sum_\mu\frac{1}{E_\mu(\beta)}\vert\psi_{R,\mu}(\beta)\rangle\langle\psi_{L,\mu}(\beta)\vert h_+$, with 
$q_\mu(\beta)$ corresponds to band $E_\mu(\beta)$, respectively. Via the $Q$ matrix,
the total winding number can be defined. For instance, in a one-dimensional (1D) system with single GBZ $\beta_{GBZ}$ \cite{ghatak2019new}
\begin{align}
    \omega_\mu=\frac{1}{2\pi i}\oint_{\beta_{GBZ}}\text{Tr}[q(\beta)^{-1}dq(\beta)]=\frac{1}{2\pi i}\oint_{\beta_{GBZ}}d\ln{\text{det}\left[q(\beta)\right]}. \label{eq:h12}
\end{align}

From Eq. (\ref{eq:Heff}), it is easy to know that in $H_{eff}$, $h_+=t_1+t_2/\beta+\gamma/2$, $h_-=t_1+t_2\beta-\gamma/2$, and the total winding number is defined as
\begin{align}
    \omega_{NH}=\frac{1}{2\pi i}\oint_{\beta_{NH}}(\frac{t_2}{2\beta(\beta(\gamma/2+t_1)+t_2)}-\frac{t_2}{2(\gamma/2-(t_1+\beta t_2))})d\beta,\label{eq:omega}
\end{align}
here the GBZ region $\beta_{NH}$ is depicted by Eq. (\ref{eq:gbz}). In Fig. \ref{Fig:wNH}, we plot the chiral winding number that changes with $t_1$. The transition points appear at $\pm\sqrt{t_2^2+(\gamma/2)^2}$, coincidence with the zero modes of the energy spectrum \cite{yao2018edge}. The conventional formulations, conversely, are usually proved to be inadequate due to the breakdown of BZ.

\subsection{The open description without jump}\label{sec:topo_withoutjump}

From Sec. \ref{sec:symmetry}, we have known that when there is no jump, the shape matrix $\textbf{A}^\prime$ (Eq. (\ref{eq:Aprime})) also exhibits chiral symmetry. To demonstrate the equivalence and validity of our scheme, we re-examine the above issue under the framework of the open system description. From Eq. (\ref{eq:pauliAk}), we have
 \begin{align} &h_\pm=-i(t_1+t_2\cos{k})(\sigma_y\otimes I_2)\mp t_2\sin{k}(\sigma_y\otimes I_2)\mp
  i\gamma I_2\otimes\sigma_z.\label{eq:openh12}
\end{align}

In this case, calculating the winding number of the system will be a bit complicated, since we will face the problem of integration over multiple energy bands $E_\mu$ and their corresponding $\text{GBZ}_\mu$. 
By referring Eq. (\ref{eq:A0})-(\ref{eq:Ak}), we can write out 
the eigenvalues $\lambda\in\cup\{E_\mu\}$ of $\textbf{A}(\beta)$ as  (eigenvalues that satisfy $\text{Re} \lambda\geq0$ are usually referred to as ``\emph{rapidities}'' \cite{prosen2008third})
\begin{align}
E_{1,\pm}(\beta)&=\gamma/4\pm\sqrt{(\gamma/2+t_1+t_2\beta)(\gamma/2-t_1-t_2/\beta)}/2,\nonumber\\
E_{2,\pm}(\beta)&=-\gamma/4\pm\sqrt{(\gamma/2+t_1+t_2/\beta)(\gamma/2-t_1-t_2\beta)}/2,\label{eq:Eband}
\end{align}
it should be noted that the presence or absence of jump terms will not alter the spectrum of the shape matrix because the above equation does not depend on the parameter $\kappa$. The reasons behind this characteristic and its associated properties will be explained in detail later.

For such a multi-GBZ system, one might intuitively express the total winding number as the summation of integrals over all GBZ regions $\beta_{GBZ,\mu}$
\begin{align}
\omega_{no jump}&=\frac{1}{2\pi i}\sum_\mu\oint_{\beta_{GBZ\mu}}\text{Tr}\left[q_\mu^{-1}\partial_\beta q_\mu\right],
\label{eq:omeganojump}
\end{align}
here $\beta_{GBZ,1,2}$ is determined by Eq. (\ref{eq:aGBZ1}), Eq. (\ref{eq:aGBZ2}), as shown in Sec. \ref{sec:aGBZ_nojump}.
However, this is not the case in reality, because different bands $E_\mu(\beta)$ may have distinct projections on the complex plane,
corresponding to different sub-GBZs. Moreover, their contributions are also nonadditive in nature, because
\begin{align}
  \sum_\mu\text{Tr}(q_\mu^{-1}dq_\mu)=\sum_\mu d\ln\text{det}(q_\mu)\neq d\ln\text{det}(\sum_\mu q_\mu),
  \end{align}
  thus when $\beta_{GBZ,1}=\beta_{GBZ,2}$, Eq. (\ref{eq:omeganojump}) cannot be reduced to the single GBZ case as in Eq. (\ref{eq:h12}). To solve this problem, we use the ``wave function'' method introduced in Ref. \cite{yang2020nonhermitiana}, which is defined as
  \begin{align}
      \omega_{nojump}=\frac{1}{2}(\omega_+-\omega_-),\quad \omega_{\pm}=-P_{\pm}+\sum_\mu Z_{\pm,\mu},\label{eq:cal_omega}
  \end{align}
  here $P_{\pm}$ is the order of the pole of $\text{det}(h_{\pm})$, $Z_{\pm,\mu}$ is the number of the zeros that not only satisfy $\text{det}(h_\pm(\beta))=E_\mu(\beta)$, but also inside its corresponding $\text{sub-GBZ}_\mu$. The basic idea of this method is that the poles in the interior of non-corresponding sub-GBZs will not affect the total winding number, with its proof relying on the \emph{Cauchy's argument principle} \cite{churchill2014ebook}.
 Recall the conclusions in Eq. (\ref{eq:openh12}), we will have $\text{det}(h_+)=(\gamma^2-4(t_1-it_2\beta)^2)^2/16, \text{det}(h_-)=(\beta^2(\gamma^2-4t_1^2)-8i\beta t_1t_2+4t_2^2)^2/(16\beta^4)$. Here it is worth noting that the zeros are all double degenerate, and $P_+=0,P_-=4$. In Fig. \ref{Fig:multi1},\ref{Fig:multi2}, we denote all the zeros $Z_{\pm}$ with blue $(+)$ and green $(-)$ dots. For a detailed examination, we first concentrate on Fig. \ref{Fig:multi1}. From the definition of $Z_{\pm}$, it is easy to find the zeros $\pm 2(1)$ belong to $\text{GBZ}_{1(2)}$. Since the zeros $+1,+2$ are outside $\text{GBZ}_{2,1}$, they do not contribute to the total winding number, which means $Z_{+1,2}=0$. As for the zeros $-2,+1$, they each contribute $Z_{-2,1}=2$. The above analysis will lead to a total winding number $\omega_{nojump}=0$. Likewise, in Fig. \ref{Fig:multi2}, we will have $Z_{+1,+2}=2,Z_{-1,-2}=0$, and the total winding number will take $\omega_{nojump}=4$. 
 In Fig. \ref{Fig:windingnojump}, the invariant that changes as a function with $t_1$ is shown (cyan solid line).
Furthermore, for comparison, we also re-examined the problem with BZ, whose winding number could be defined analogously as \cite{dangel2018topological}
\begin{align}
\omega_{nojump}=\frac{1}{2\pi i}\int_{-\pi}^\pi\langle l^\prime_\mu\vert\partial_k\vert r^\prime_\mu\rangle dk,\label{eq:omega_njBZ}
\end{align}
describes by the right and left eigenvectors
 $\vert r^\prime_\mu\rangle, \vert l^\prime_\mu\rangle$ of $\textbf{A}^\prime(k)$. As depicted in Fig. \ref{Fig:windingnojump} (red-dashed line), the winding number displays fractional quantization. Nevertheless, it is also evident that it fails to faithfully predict the phase transition points.
\begin{figure}[b]
\centering
\subfigure{
\label{Fig:jumpm2}
\includegraphics[width=0.24\textwidth]{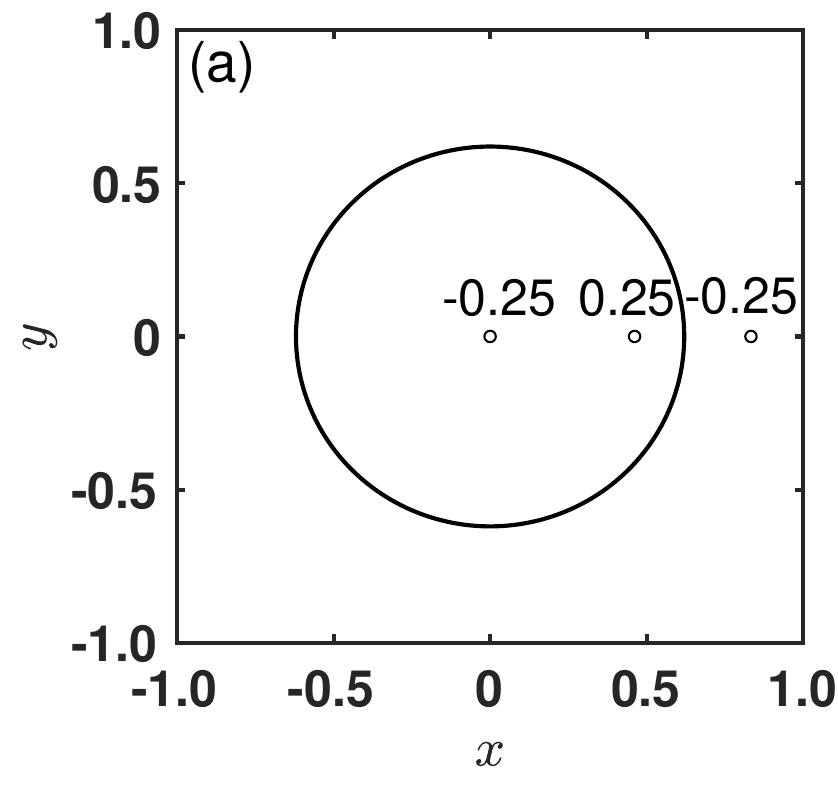}}\subfigure{
\label{Fig:jumpm1}
\includegraphics[width=0.24\textwidth]{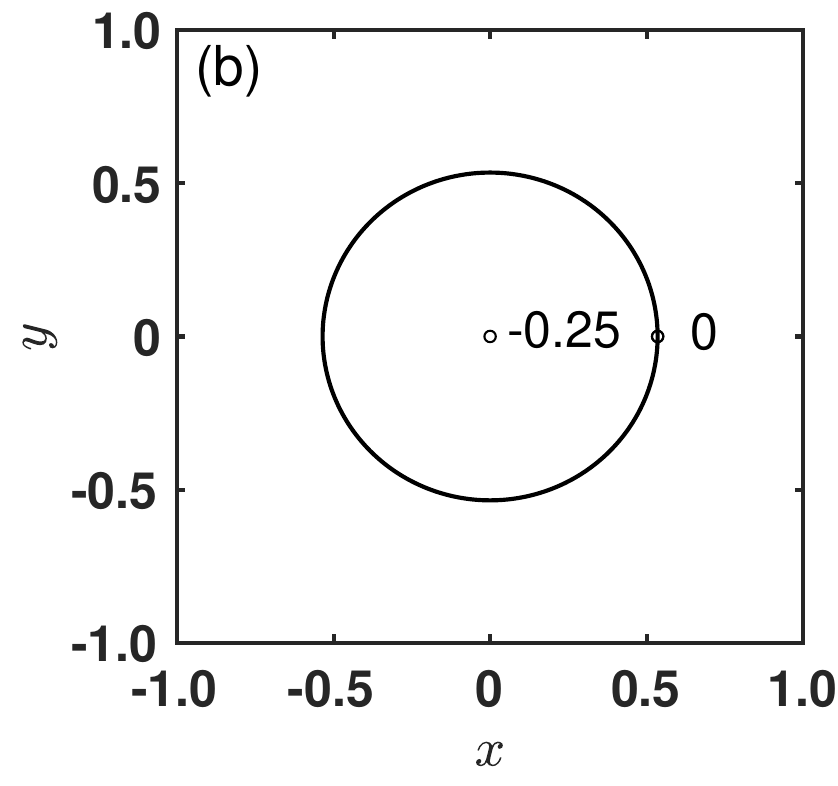}}\subfigure{
\label{Fig:jumpp1}
\includegraphics[width=0.24\textwidth]{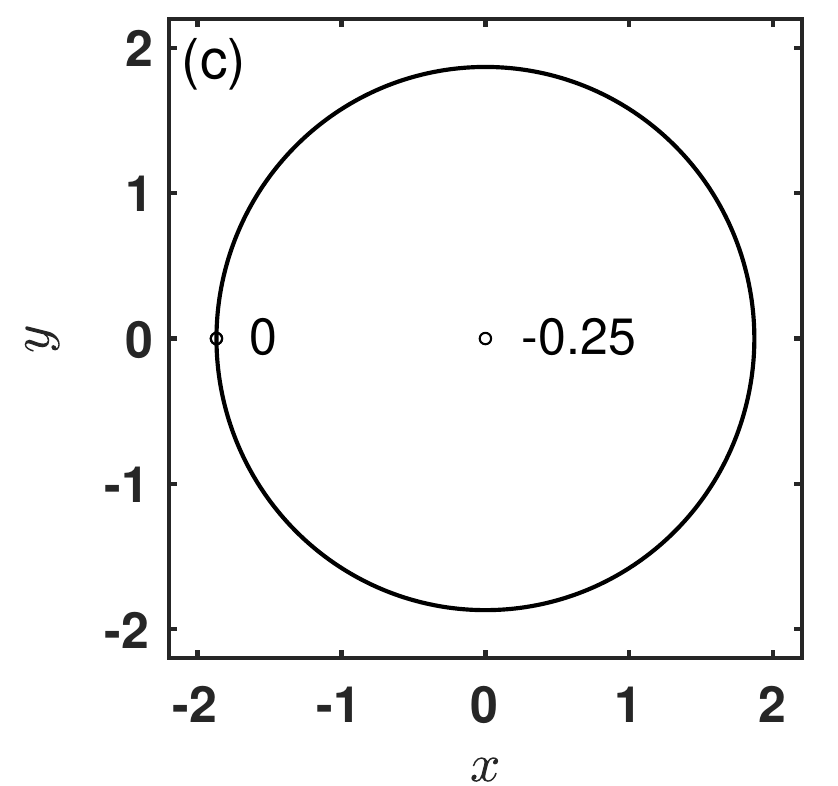}}\subfigure{\label{Fig:jumpp2}
\includegraphics[width=0.24\textwidth]{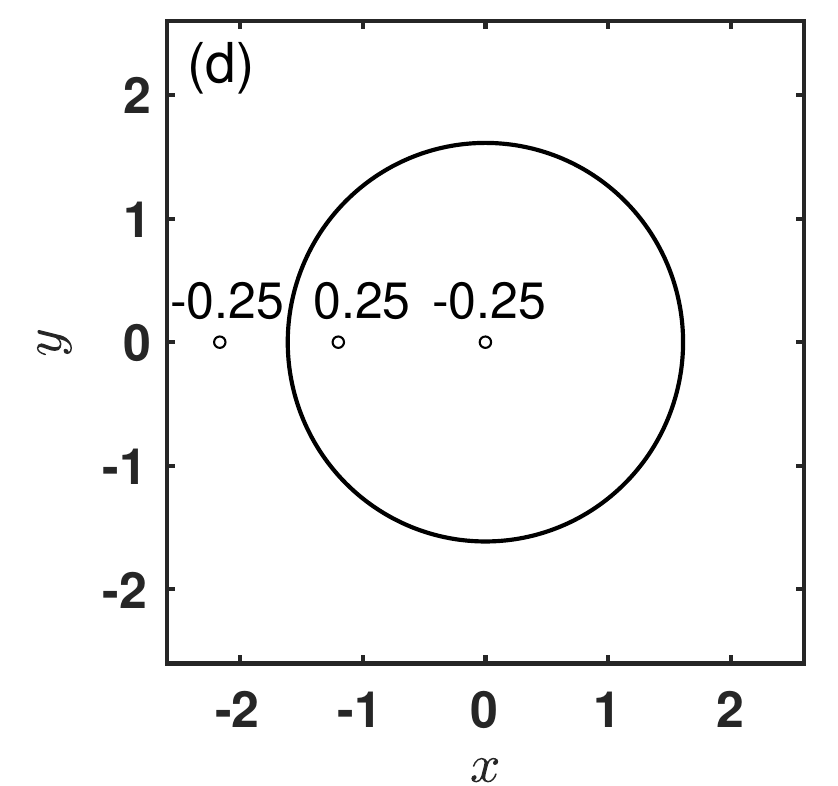}}
\subfigure{
\hspace{1.5mm}
\label{Fig:jumpphase}
\includegraphics[width=0.75\textwidth]{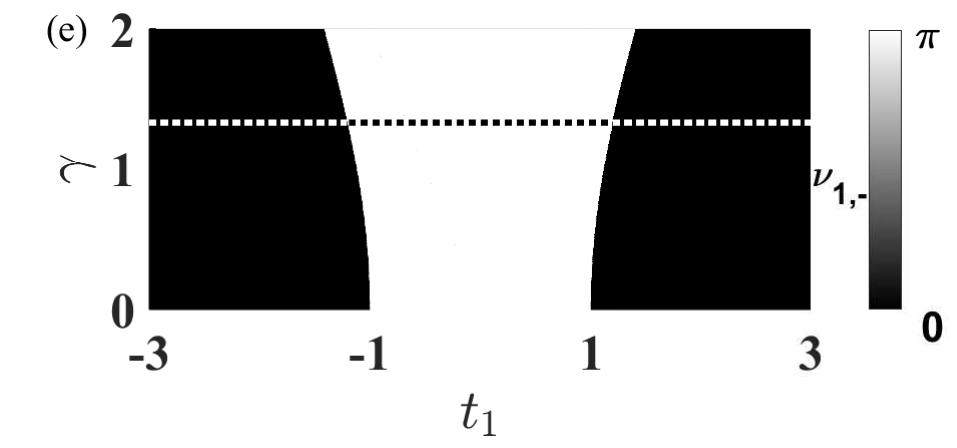}}
    \caption{(a-d) The GBZ of band $E_{1,-}$ when jumping terms are included. Parameters are chosen as (a) $t_1=-1.5$, (b) $t_1=-\sqrt{(\gamma/2)^2+t_2^2}$, (c) $t_1=\sqrt{(\gamma/2)^2+t_2^2}$, (d) $t_1=1.5$. The numbers are the residues of the corresponding pole of Zak phase $\nu_{1,-}$. The $0$ residue in (b)(c) is the second-order pole, while others are of the first order.  (e) The phase diagram of Zak phase $\nu_{1,-}$ respect with $t_1$ and $\gamma$, where $\kappa$ takes $1$. The black region corresponds to a $0$ Zak phase, while the white region represents a $\pi$ Zak phase. The dashed line represents the previous set of parameters when $\gamma=4/3$.}
\end{figure}

Through the example, it is clear that our scheme can predict phase transition points as in NH theory, since the information is encoded within the shape matrix. A comparative analysis between Fig. \ref{Fig:wNH} and Fig. \ref{Fig:windingnojump} reveals an intriguing alignment: despite the disparities in dimensionality and descriptive frameworks employed in these two contexts, the transition points appear at the same location. This congruence provides compelling evidence regarding the efficacy of our approach in characterizing phases within open quantum systems.

\subsection{The jump involved case}\label{sec:topo_withjump}

By extending the process to the entire open system, we utilize the concept of the Zak phase to characterize its topology. In 1D system, it is regarded as an analog of the Berry phase that the particle picks up when moving across the BZ \cite{Zak1989berry}, and has been widely applied in dissipative systems and NH systems \cite{garrison1988complex,hatsugai2006quantized}.

In Sec. \ref{sec:symmetry}, we discussed how the presence of jumping terms reveals the time-reversal symmetry of the shape matrix $\textbf{A}$, described by the relation $\textbf{A}(-k)=U_T\textbf{A}^*(k)U_T^\dagger$. Here, $U_T=I_2\otimes\sigma_y\otimes\sigma_x$ is a unitary and Hermitian matrix. We can judiciously set $\mathcal{P}=U_T$ and $\mathcal{T}=\mathcal{K}$, with $\mathcal{K}$ being the conjugation operator. Consequently, $\textbf{A}(k)\mathcal{P}\mathcal{T}=\mathcal{P}\mathcal{T}\textbf{A}(k)$ holds, which reveals that the shape matrix remains invariant under the combined action of parity and time inversion, expressed as $\left[\textbf{A},\mathcal{P}\mathcal{T}\right]=0$. Following this analysis, the Zak phase for a given band $\mu$ can be defined as
\begin{align}
\nu_{\mu}=i\oint_{\beta_{{GBZ}_\mu}}\langle l_{\mu}\vert\partial_{\beta}\vert r_{\mu}\rangle d\beta,\label{eq:Zak}
\end{align}
where $\beta_{{GBZ}_\mu}$ is the GBZ of the selected band, the $\vert l_\mu\rangle$ and $\vert r_\mu\rangle$ are the left and right eigenvectors of $\textbf{A}(\beta)$, corresponding to the band $E_\mu$. It is worth to stress that the derivation of $\beta$ is an extension of Hermitian cases ($k$).

As an example, in Fig. \ref{Fig:jumpm2}-\ref{Fig:jumpp2}, we plot the GBZ of the band $E_{1,-}$ for several sets of parameters, when jump is presented ($\kappa=1$). By using Eq. (\ref{eq:Zak}), one can derive the analytical expression for the Zak phase associated with band $E_{1,-}$
\begin{align} \nu_{1,-}=i\oint_{\beta_{GBZ}}\frac{t_2(2\beta^2(t_1-\gamma/2)+4\beta t_2+2t_1+\gamma)}{2\beta(\beta(\gamma-2t_1)-2t_2)(\gamma+2(t_1+\beta t_2))},
\end{align}
where the GBZ region is the loop described by Eq. (\ref{eq:aGBZ1}), corresponding to the band $E_{1,-}$. In this way, the Zak phases for all the bands can be calculated.

In Fig. \ref{Fig:jumpphase}, we display the phase diagram illustrating the dependence of the Zak phase $\nu_{1,-}$ on the parameters $t_1$ and $\gamma$. In full area the Zak phase is quantized, where the black and white region contribute a $0$ or $\pi$ phase, respectively. The dashed line stands for $\gamma=4/3$ in Fig. \ref{Fig:jumpm2}-\ref{Fig:jumpp2}.

\begin{figure}[htbp]
\centering
\includegraphics[width=0.8\textwidth]{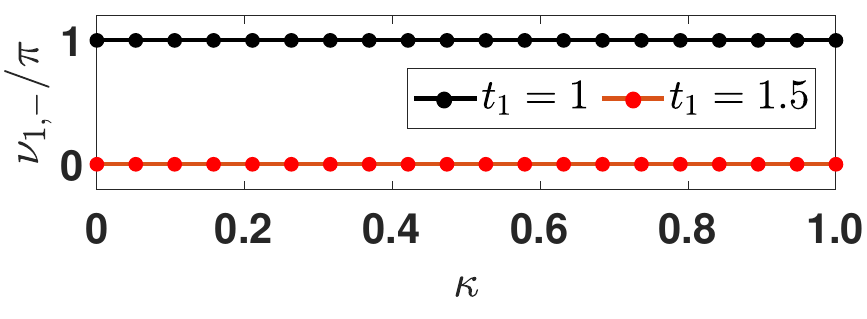}
    \caption{The numerical result of the Zak phase $\nu_{1,-}$ by using Eq. (\ref{eq:Zak}), corresponding to $E_{1,-}$ band, which respects to different jump parameters $\kappa$. Parameters are chosen as $t_2=1, \gamma=4/3$. The black line and dots represent $t_1=1$ case, while the red line and dots represent $t_1=1.5$. The Zak phase remains unchanged under different values of $\kappa$.}\label{Fig:kappa}
\end{figure}

\begin{figure}
\centering
\includegraphics[width=0.6\textwidth]{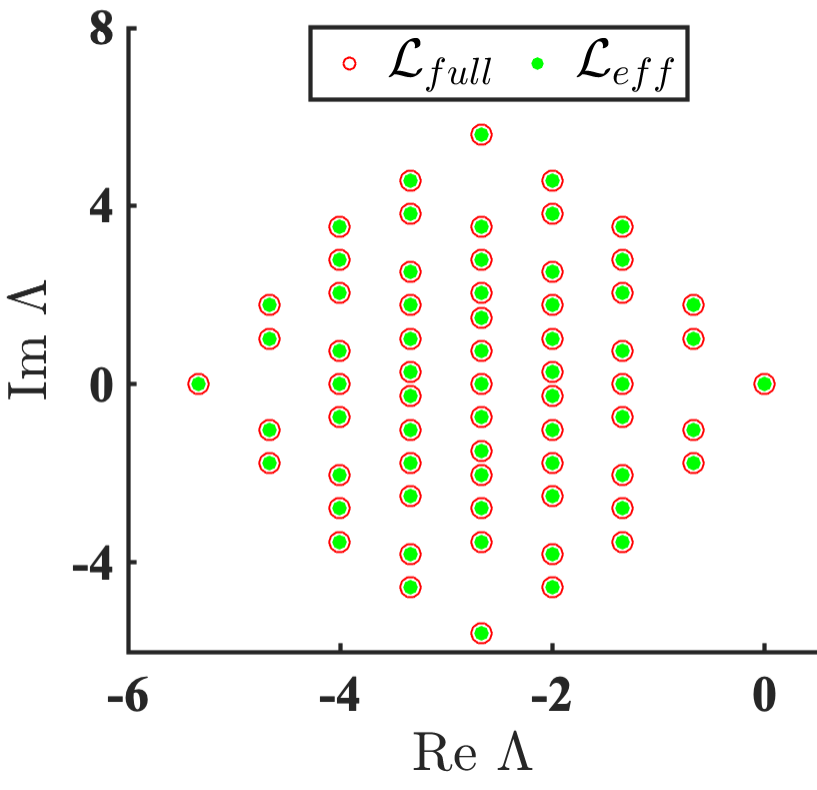}
    \caption{Liouvillian spectrum $\Lambda$ of the pure loss model under OBC. The red cycles represent the $\mathcal{L}_{full}$ and the green dots are for $\mathcal{L}_{withoutjump}$. We restrict $N=4$, and the other parameters are chosen as $t_1=1.5,t_2=1,\gamma=4/3,\kappa=1$. This result can also obtained by the exact diagonalization method in Ref. \cite{zhangExactDiagonalizationBose2010}.}\label{Fig:Liou}
\end{figure}

Finally, to visualize the effect of quantum jump events, in Fig. \ref{Fig:kappa}, we numerically calculate $\nu_{1,-}$ for various parameters of $\kappa$ and $t_1$. Intriguingly, it indicates that in this model, the strength of jump $\kappa$ does not influence the exhibited Zak phase. However, we shall emphasize this observation is a model-dependent issue.  In the context of the pure loss condition, when writing the Liouvillian operator in the matrix representation, the jumping terms $\mathcal{L}_{jump}$ will link the subspaces labeled by $(N,N-1)$, where $N$ represents the total number of fermions. Conversely, the other components, $\mathcal{L}_{withoutjump}=\mathcal{L}_0+\mathcal{L}_{eff}$, conserve the number of particles, which links the subspaces of $(N,N)$. 
Mathematically, when in the absence of the jumping terms, the Liouvillian can be represented into  a block-diagonal form, whereas the presence of jumping terms will result in a block-upper-triangular form.
In accordance with the properties of determinants, the upper triangular part has no influence on the Liouvillian spectrum $\{\Lambda\}$, regardless of PBC or OBC \cite{yoshida2020fate,niu2023effect}.
Furthermore, such eigenvalue structure will also imply that their rapidities are identical \footnote{Actually, we have alreadly shown that for \text{PBC}, straightforward calculation immediately reveals that $\textbf{A}$ and $\textbf{A}^\prime$ have the same energy spectra (independent of $\kappa$).}, as the Liouvillian spectrum ${\Lambda}$ is in fact a linear combination of the rapidities $\{\lambda_j\}$ \cite{prosen2008third}
\begin{align}
\Lambda_{\underline{\nu}}=-2\sum_j\lambda_j\nu_j,
\end{align}
where $\underline{\nu}=(\nu_1,\nu_2,\dots,\nu_{2n})$ and $\nu_j\in\{0,1\}.$ In Fig. \ref{Fig:Liou}, we plot the full Liouvillian spectrum $\mathcal{L}_{full}=\mathcal{L}_{eff}+\kappa\mathcal{L}_{jump}$ (red circles) and jump absence spectrum $\mathcal{L}_{eff}$ (green dots).  Notably, their spectra show no difference, which proves our conjecture.
According to the characteristic equation Eq. (\ref{eq:Djump1}),(\ref{eq:Djump2}), the same spectrum induces the same region of GBZ, regardless the value of $\kappa$.

In order to elucidate the effect of quantum jump, we now focus on the circumstance where both loss and gain are incorporated into the Lindblad operators as in Eq. (\ref{eq:loss}), (\ref{eq:gain}). In fact, considering the presence of gain is imperative, as it forms the foundation for achieving a non-trivial NESS.
In such situations, it is no longer possible to represent the jumping terms into the upper triangular block form, which may indicate that they may be non-negligible.

For the sake of simplicity, we reconsider the sub-GBZ ($\kappa=1$) when the system exhibits both loss and gain. In this case, the sub-GBZ remains circular, with a radius $r$ of
\begin{align}
    r=\left|\frac{\gamma_l/2-\gamma_g/2+t_1}{\gamma_l/2-\gamma_g/2-t_1}\right|^{\pm\frac{1}{2}}.\label{eq:radius}
\end{align}

At this juncture, the radius of GBZ relies on the difference between loss and gain rate $\gamma_l-\gamma_g$, rather than their summation $\gamma_l+\gamma_g$ \footnote{As in Sec. \ref{sec:GBZ_NH}, by taking $\gamma\rightarrow\gamma_l+\gamma_g$, apparently the radius of \text{GBZ} depends on the summation of $\gamma_l$ and $\gamma_g$.}. Furthermore, when $\kappa$ takes other values, the GBZ may deviate from being a perfect circle. In these cases, the jumping terms can not be safely neglected, which differs from the circumstance when there is only loss.

\begin{figure}[htbp]
\centering
\includegraphics[width=0.7\textwidth]{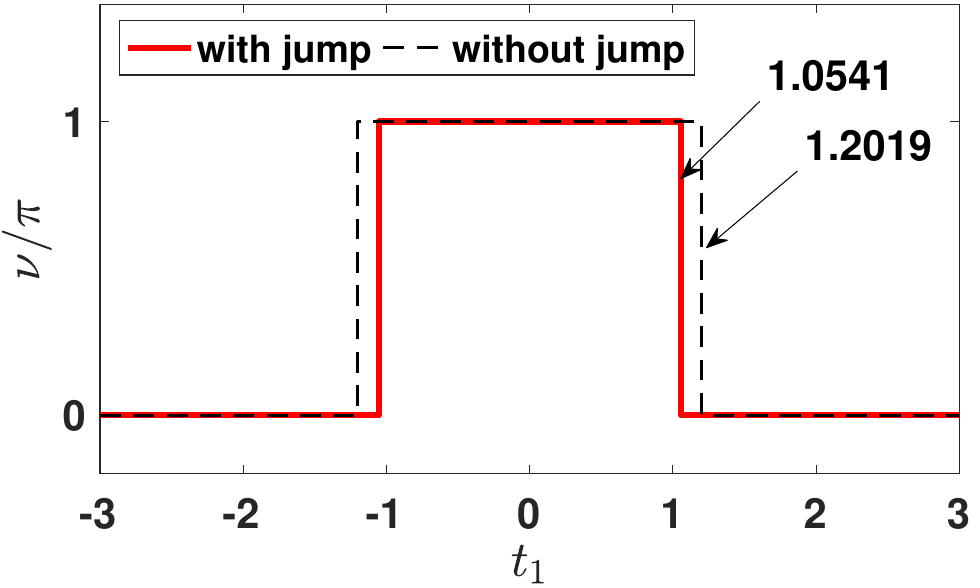}
    \caption{The Zak phase $\nu$ of a selected band when the system is suffering both loss and gain ($\kappa=1$, red solid line), and the jumping absence case ($\kappa=0$, black dashed line, as discussed in the main text), changes with $t_1$. The transition points changed, which indicates that the jumping terms can not be safely neglected. Parameters are chosen as $t_2=1,\gamma_l=1,\gamma_g=1/3.$}\label{Fig:gainZak}
\end{figure}

In Fig. \ref{Fig:gainZak}, we illustrate the Zak phase when the system is subject to loss and gain simultaneously (red solid line).
Additionally, for comparison, we also provide a plot of the results when the jumping terms are entirely neglected (black dashed line). The transition points are at $\pm\sqrt{t_2^2+((\gamma_l-\gamma_g)/2)^2}$, which is different from the no jumping case. This also demonstrates the necessity of our approach.

To summarize, the effect of quantum jumping events in the system needs to be carefully examined and may significantly affect the conclusions in effective NH theory. In light of this, our proposed approach has demonstrated its effectiveness and can be instrumental in tackling this issue.

\subsection{Dynamical signatures}\label{sec:dynamical_signatures}

In this subsection, we will provide an explanation for the distinctive phase transitions observed in the previous subsection, which differs from the effective NH theory $H_{eff}$. In a sense, these transitions can be understood from the perspective of dynamical signatures in transient dynamics, which have been demonstrated to play an essential role in the topology of open systems \cite{vancaspel2019dynamical,kawasaki2022topological}. To be specifically, the local evolution of observables, can be attributed to the locality arising from the edge master mode. This can be illustrated by a straightforward example, the one-site occupation number $C_{ij}(t)=\text{Tr}(\rho(t)c^\dagger_ic_j)$, whose evolution obeys the following equation \cite{song2019nonhermitiana}
\begin{align}
\frac{dC(t)}{dt}=i\left[h^T,C(t)\right]-\{M_l^T+M_g,C(t)\}+2M_g,
\end{align}
where $M_l=\sum_\mu D^{l*}_{\mu i}D^{l}_{\mu j}$, $M_g=\sum_\mu D^{g*}_{\mu i}D^{g}_{\mu j}$ is the Hermitian matrices that parametrizing the Lindblad operators $L_\mu$.  Therefore, the deviation of the dynamics from the steady state $C(\infty)=C_s$ can be expressed as $\tilde{C}(t)=C(t)-C_s$, which satisfies \cite{song2019nonhermitiana}
\begin{align}
\tilde{C}(t)=e^{Xt}\tilde{C}(0)e^{X^\dagger t},
\end{align}
here $X=ih^T-(M_l^T+M_g)$ is the damping matrix.
For the sake of simplicity, we start from the periodic boundary chain. Apparently, in the $k$ space $h(k)=(t_1+t_2\cos k)\sigma_x+t_2\sin{k}\sigma_y$, and $M_l(k)=\gamma_l(I-\sigma_y)/2,M_g(k)=\gamma_g(I-\sigma_y)/2$, hence $X(k)$ writes \cite{song2019nonhermitiana}
\begin{align}
&X(k)=ih^T(-k)-M^T_l(-k)-M_g(k)\nonumber\\&=i((t_1+t_2\cos{k})\sigma_x+(t_2\sin{k}+\frac{i}{2}(\gamma_l-\gamma_g))\sigma_y)-\frac{\gamma_l+\gamma_g}{2}I,
\end{align}
which resembles a NH SSH model, except that its non-reciprocal hopping strength becomes $(\gamma_l-\gamma_g)/2$. A similar understanding could also be found in Ref. \cite{mcdonald2022nonequilibrium}. Recall the conclusions from Sec. \ref{sec:topo_NH}, it is not difficult to observe that they align with the findings discussed in the previous subsection \footnote{The phase transition points can be obtained via a variable substitution, replacing $\gamma$ with $\gamma_l-\gamma_g$.}.
These correspondences strengthen the validity of our findings.

\subsection{BBC}\label{sec:BBC}

\begin{figure}
\centering
\subfigure{
\label{Fig:ReA}
\includegraphics[width=0.48\textwidth]{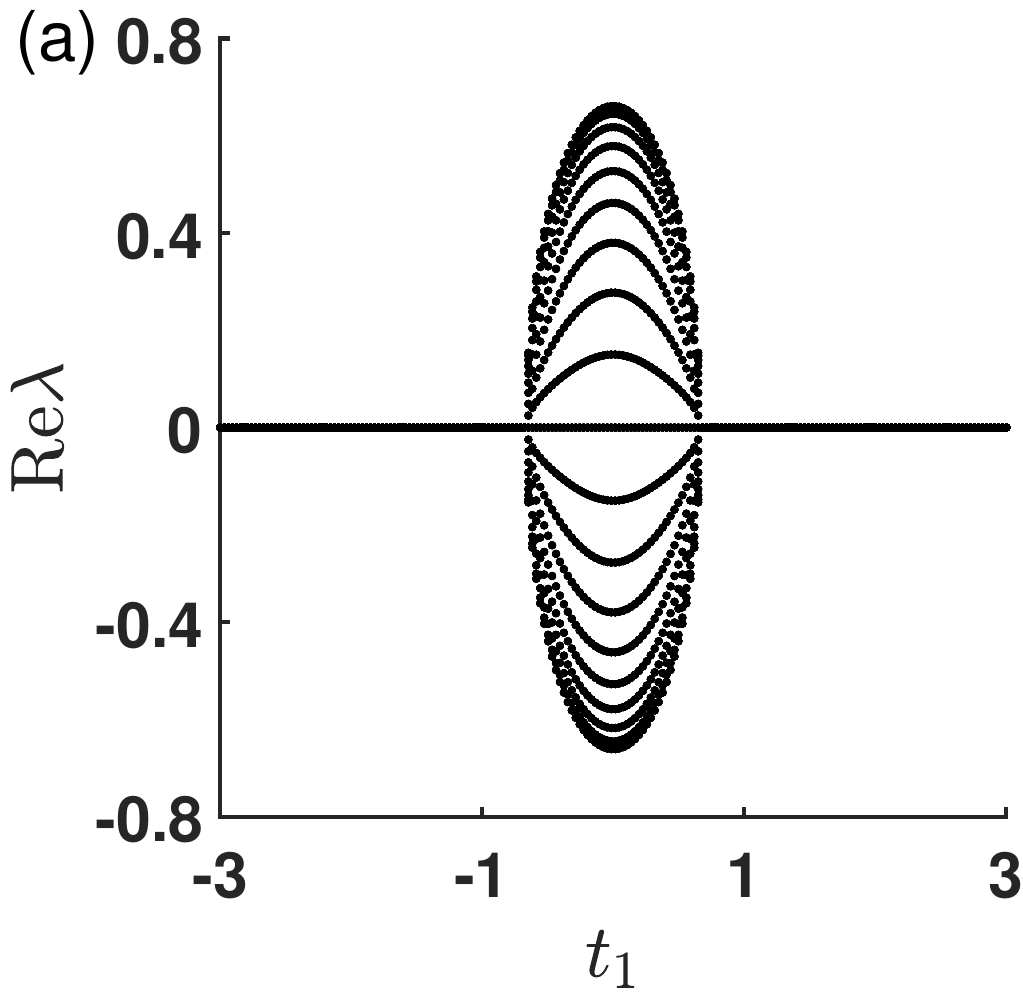}}\subfigure{
\label{Fig:ImA}
\includegraphics[width=0.48\textwidth]{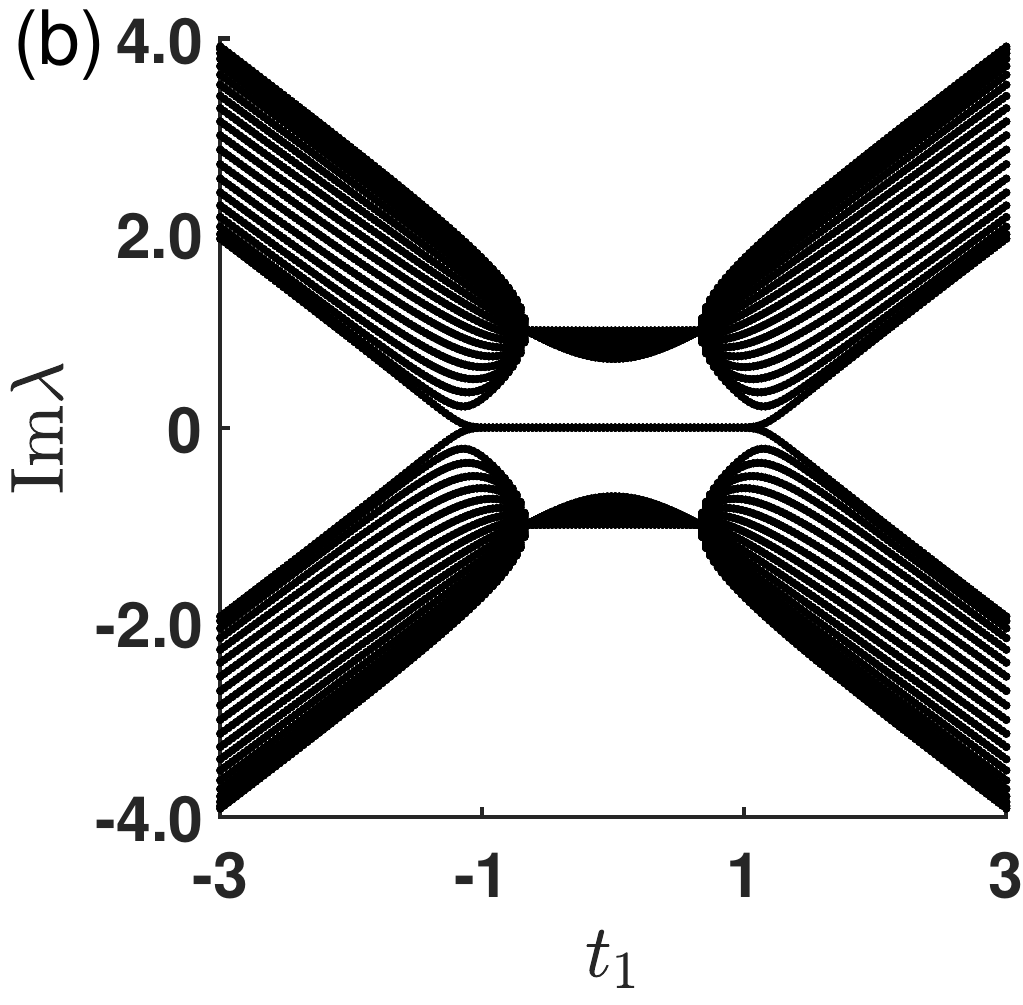}}
    \caption{(a-b) The real (a) and imaginary part (b) of numerical spectrum $\{\lambda\}$ of the shape matrix $\textbf{A}^\prime$ with an open chain, here we choose $L=20$ (unit cell), $t_2=1$, and $\gamma=4/3$, $t_1$ varies from $-3$ to $3$.}
\end{figure}

In this subsection, we will show that the BBC still holds in our scheme. Here, for simplicity, we mainly talk about the purely loss scenario of $\textbf{A}^\prime$ 
as an example. In Fig. \ref{Fig:ReA}-\ref{Fig:ImA}, we numerically calculate the eigenvalues $\{\lambda\}$ of the shape matrix $\textbf{A}^\prime$ under different parameters of $t_1$.
Following the insights provided by Fig. \ref{Fig:spectraB},\ref{Fig:ImA}, we can see the edge states located at $\text{Im}\ \lambda=0$ (eightfold degenerate). By contrast, from Fig. \ref{Fig:spectraA}, there shows no edge states, because the model experiences a phase transition at $t_1=\sqrt{(\gamma/2)^2+t_2^2}$. In Fig. \ref{Fig:windingnojump}, we can observe a one-to-one correspondence between the invariants and the boundary states, which indicates there exist a winding number with 0 or 4. Apparently, it matches the number of boundary states (0 or 8), because we have boundary states at each ends of the system. For $\textbf{A}$, by using Eq. (\ref{eq:Zak}), the total Zak phase of the system $\sum_\mu\nu_{\mu}$ is 0 or 8$\pi$, thus the same conclusion holds. In summary, the BBC of the model still preserves, which proves the validity and feasibility of our scheme.

\section{Conclusion\label{sec:conclusion}}

In recent years, NH systems have been regarded as the cherished tools in dissipative modeling. As a primary distinction compared to open systems, the role of quantum jumping terms may be of paramount importance. Nonetheless, to our knowledge, the repercussions of omitting the jumping terms have not been adequately explored. In this study, we investigate the similarities and differences between these two approaches from a topological perspective. With an illustrative SSH model which subjects to loss and gain, we demonstrate that the topology of open systems can be effectively encoded within their corresponding shape matrix. When in the absence of jump, there are considerable similarities with the traditional NH topology in terms of symmetry, transition points, BBC, etc. Due to the dynamical signatures in transient dynamics, the introduction of quantum jumps may alter the system's topology. These indications suggest that jumps may play an irreplaceable role in dissipative processes, which is beyond the scope of the NH theory. Our approach establishes a connection between NH and open system representations, and sheds light on the unique role of quantum jumps.

\section*{Acknowledgments}
We thank Professor X.X.Yi for his helpful discussions. This work was supported by the National Natural Science Foundation of China (NSFC) under Grants No. 12175033, No. 12147206 and National Key R$\&$D Program of China (No. 2021YFE0193500).

\appendix

\section{THE CALCULATION OF GENERALIZED BRILLOUIN ZONE \label{app:gbz}} The calculation of the generalized Brillouin zone in the main text is basically based on the scheme proposed in Ref. \cite{yao2018edge,yokomizo2019nonBloch}, where the usually used BZ in Hermitian case can be considered as a specific circumstance. Here we give out a brief procedure for the calculation of GBZ.

(i) Suppose a 1D tight-binding model with total dimension $2M$, the Hamiltonian is given by
\begin{align}
    H=\sum_{n,i,\mu,\nu}t_{i,\mu\nu}c^\dagger_{n+i,\mu}c_{n,\nu},
\end{align}
where $n$ represents the $n^{\text{th}}$ cell of the model, $\mu,\nu$ is the index of the fermion in an individual cell, which contains multiple degrees like momentum, spin, etc. $t$ is the hopping strength and different $i$ indicates the long-range interactions of lattices in different cells.

(ii) The eigen-function of Hamiltonian satisfies $H\vert\Psi\rangle=E\vert\Psi\rangle$,  the eigenvector $\vert\Psi\rangle$ in the real space can be written as a linear combination
\begin{align}
    \vert\Psi\rangle=(\dots,\psi_{n,1},\psi_{n,2},\dots,\psi_{n+1,1},\dots)^T,\label{eq:psi}\\
    \psi_{n,\mu}=\sum_{j} \phi^{(j)}_{n,\mu}, \phi^{(j)}_{n,\mu}=(\beta_j)^{n}\phi_{u}^{(j)},\label{eq:sumpsi}
\end{align}
where the second line is because of spatial periodicity.

(iii) After simplification, the
 characteristic equation of the Bloch form Hamiltonian $\mathcal{H}(\beta)$ can be presented as
 \begin{align}
     f(\beta,E)=\text{det}(\mathcal{H}(\beta)-E)=\frac{P(\beta,E)}{\beta^p}=0,\label{eq:character}
 \end{align}
 where $\mathcal{H}(k)$ is performed by a substitution $e^{ik}\rightarrow \beta$. The norm of $\beta$ is not necessarily unity, which is the reason why the ordinary BZ breaks down in NH cases. The $p$ is the order of the pole of $P(\beta,E)$, and $f(\beta,E)$ can be trivially factorized as
\begin{align}
f(\beta,E)=\prod_\mu^n(E-E_\mu(\beta))=0, \label{eq:multiband}
\end{align}
for multi bands, here $n$ is the dimension of Bloch Hamiltonian $\mathcal{H}$.

 (iv) Numerically calculate the eigenvalues $E_i$ $(i=1,2,\dots 2M)$ of $H$, and the $\beta$ can be solved by function
\begin{align}
f(\beta,E_i)=0,
\end{align}
since $f(\beta,E)$ can be expressed as a $2M$-order polynomial in terms of $\beta$,
apparently, there are $2M$ solutions of $\beta$. To pick out the values of $\beta$, an important feature is that for bulk-band property, for the solution labeled as $\vert\beta_1\vert\leq\vert\beta_2\vert\leq\dots\leq\vert\beta_{2M}\vert$, the continuum bands requires a pair of $\beta$ that satisfy
\begin{align}
    \vert\beta_M\vert=\vert\beta_{M+1}\vert,\label{eq:equalbeta}
\end{align}
which determines the trajectory of GBZ. In other words, there must exist a pair of conjugate points $\beta_M=\beta_0,\beta_{M+1}=\beta_0e^{i\theta}$, where $\beta_0$ is a complex number.  When $H$ is Hermitian, it can be easily proved that GBZ is always a unit cycle (BZ) \cite{yokomizo2019nonBloch}.
\\

\section{THE CALCULATION OF aGBZ}\label{app:aGBZ}
From \ref{app:gbz}, we have known that $f(\beta,E)=0$ and there shall be at least a pair of solutions $\beta$ which has the same modulus $\vert\beta_M\vert=\vert\beta_{M+1}\vert$. Thus we can assume $\beta_M=\beta_0$ and $\beta_{M+1}=\beta_0e^{i\theta}$, where $\beta_0$ is a complex number, which results in
\begin{align}
    f(\beta_0,E)=f(\beta_0e^{i\theta},E)=0,\theta\in\mathcal{R}. \label{eq:betatheta}
\end{align}

However, in practical applications, the numerical diagonalization of NH systems is particularly sensitive to boundaries. The minor deviations arising from the selection of model size and computational precision can have a substantial impact on the contour of the GBZ. A small system size may be insufficient to accurately depict the GBZ. Furthermore, in some cases, increasing the size may not yield substantial improvements and could, on the contrary, lead to a significant increase in computation cost and running time. An inappropriate value selection may cause unpredictable errors, or unacceptable calculation time (usually up to days).  Additionally, in subsequent calculations we will encounter the integration over the GBZ, an analytical will possess unparalleled advantages. Here, we referred to the method introduced in Ref. \cite{yang2020nonhermitiana}, where the authors employ a mathematical technique known as the \emph{Resultant} to handle this issue.

In math, the \emph{Resultant} is regarded as a utilized tool for selectively variable elimination. Suppose now there are two polynomials $f(x)=a_nx^n+\dots+a_0=\prod^n_{i=1}(x-\xi_i),g(x)=b_mx^m+\dots+b_0=\prod^m_{j=1}(x-\eta_j)$ which respect to the variable $x$, their \emph{Resultant} is defined as
\begin{align}
R_x(f,g)=a_n^mb_m^n\prod_{i,j}(\xi_i-\eta_j).\label{eq:resultant}
\end{align}

Suppose the two polynomial equations are also related to $y$, $f(x,y),g(x,y)=0$. If we want to eliminate the variable $x$ and obtain an equation containing only $y$, the following condition holds
\begin{align}
    R_x[f(x,y),g(x,y)]=0,\label{eq:eliminate}
\end{align}
we can see that this property satisfies the requirement for variable elimination $E$ in Eq. (\ref{eq:betatheta}), that is
 \begin{align}
     G(\beta,\theta)=R_E(f(\beta,E),f(\beta e^{i\theta},E))=0.\label{eq:GE}
 \end{align}

 Obviously Eq. (\ref{eq:GE}) requires both real and imaginary parts of $G(\beta,\theta)$ to be zero, which means
 $Re[G(\beta,\theta)]=0,Im[G(\beta,\theta)]=0$.
 
 Next, we would like to eliminate $\theta$ in the same fashion. However, there is a slight difference from the previous situation. The resultant operates on polynomials, however, $\theta$ appears in the exponent $e^{i\theta}$. Therefore,
 $\text{Re}G(\beta,\theta),\text{Im}G(\beta,\theta)$ are not algebraic functions of $\theta$,
 and cannot be directly eliminated in this manner. Instead, we can use the Weierstrass substitution $\cos{\theta}=(1-t^2)/(1+t^2),\sin{\theta}=2t/(1+t^2)$ for variable substitution, and $t$ can be eliminated as follows
\begin{align}
R_t[\text{Re}(G(\beta,t)),\text{Im}(G(\beta,t))]=0.\label{eq:ReImG}
\end{align}

After the above treatments, both the variables $E$ and $\theta$ will not appear anymore. However, at the same time, it may introduce trivial solutions and result in a zero value. This requires an elimination of the common trivial solutions in $\text{Re}(G(\beta,t))$ and $\text{Im}(G(\beta,t))$ through factoring. After going through the aforementioned steps, we finally obtain a nontrivial analytical expression that characterizes the curves of aGBZ. Generally, the actual GBZ is usually a subset of aGBZ.

\section{THE MULTI GBZ CASES}\label{app:multigbz}

With the help of \emph{resultant}, we obtain the aGBZ of the shape matrix $\textbf{A}^\prime$ in Sec. \ref{sec:GBZ}, the next urgent task is to verify the actual area of GBZ. In \ref{app:gbz}, we have known that the OBC spectrum $\{E\}$ is required in order to obtain the information of GBZ. However, in Eq. (\ref{eq:GE}), we have eliminated the degree $E$, which means that $E$ can take all possible values across the complex plane. Nevertheless, in practice the values that can be taken from the energy spectrum $\{E\}$ are limited, which may induce the GBZ usually being a subset of aGBZ. To address this issue, Ref. \cite{yang2020nonhermitiana} introduced the concept of a new kind of singular points known as self-conjugate points, that will only occur in NH systems. Recall that the Van Hove singularity $k_s$ in Hermitian system \cite{van1953occurrence}, which satisfies
\begin{align}
    \partial_kE_\mu(k_s)=0.
\end{align}

From the extension $k=-i\ln{\beta}$, the singularity points $\beta_0$ can be generalized to NH bands, which writes 
\begin{align}
\partial_kE_\mu(\beta_0)=0.
\end{align}

In order to search the singularity points in all bands, we can use
\begin{align}
    f(E_0,\beta_0)=0,
\end{align}
where $E_0$ satisfies
\begin{equation}
\left\{
\begin{aligned}
    f(E_0,\beta)=0,\\
    \frac{\partial f(E_0,\beta)}{\partial \beta}=0,
\end{aligned}
\right.\label{eq:cpoints}
\end{equation}
it is proved such self-conjugate points not only have $\vert\beta_M\vert=\vert\beta_{M+1}\vert$, but also $\beta_M=\beta_{M+1}$ \cite{yang2020nonhermitiana}. The contours of aGBZ that possess the self-conjugate points must form GBZ.

\nocite{*}
\newcommand{\newblock}{}
    \bibliographystyle{unsrt}
    \bibliography{niu_topo}
\end{document}